\documentclass[prd,12pt,nofootinbib]{revtex4}
\usepackage{graphicx}
\usepackage{epstopdf}
\usepackage{graphicx}
\usepackage{setspace}
\usepackage{amsmath}
\usepackage{latexsym, amssymb}
\usepackage{verbatim}
\usepackage{epstopdf}
\usepackage{lscape}
\usepackage{rotating}
\usepackage{feynmp}
\usepackage{subfloat}
\def\beq{\begin{equation}}
\def\eeq#1{\label{#1}\end{equation}}
\def\eeqn{\end{equation}}

\newenvironment{Eqnarray}%
   {\arraycolsep 0.14em\begin{eqnarray}}{\end{eqnarray}}
\def\beqa{\begin{Eqnarray}}
\def\eeqa#1{\label{#1}\end{Eqnarray}}
\def\eeqan{\end{Eqnarray}}
\def\CR{\nonumber \\ }

\def\leqn#1{(\ref{#1})}

\let\bar=\overbar

\def\bra#1{\left\langle{ #1} \right|}
\def\ket#1{\left| {#1} \right\rangle}

\def\lsim{\mathrel{\raise.3ex\hbox{$<$\kern-.75em\lower1ex\hbox{$\sim$}}}}
\def\gsim{\mathrel{\raise.3ex\hbox{$>$\kern-.75em\lower1ex\hbox{$\sim$}}}}

\def\Im{{\rm Im}}

\def\M{{\cal M}}

\def\One{{\bf 1}}

\def\del{\partial}
\def\Dslash{\not{\hbox{\kern-4pt $D$}}}
\def\dslash{\not{\hbox{\kern-2pt $\del$}}}

\def\msb{{\bar{\scriptsize M \kern -1pt S}}}

\def\drb{{\bar{\scriptsize D \kern -1pt R}}}

\makeatletter
\def\section{\@startsection{section}{0}{\z@}{5.5ex plus .5ex minus
 1.5ex}{2.3ex plus .2ex}{\large\bf}}
\def\subsection{\@startsection{subsection}{1}{\z@}{3.5ex plus .5ex minus
 1.5ex}{1.3ex plus .2ex}{\normalsize\bf}}
\def\subsubsection{\@startsection{subsubsection}{2}{\z@}{-3.5ex plus
-1ex minus  -.2ex}{2.3ex plus .2ex}{\normalsize\sl}}

\renewcommand{\@makecaption}[2]{%
   \vskip 10pt
   \setbox\@tempboxa\hbox{\small #1: #2}
   \ifdim \wd\@tempboxa >\hsize     
       \small #1: #2\par          
     \else                        
       \hbox to\hsize{\hfil\box\@tempboxa\hfil}
   \fi}

 \def\citenum#1{{\def\@cite##1##2{##1}\cite{#1}}}
 
\newcount\@tempcntc
\def\@citex[#1]#2{\if@filesw\immediate\write\@auxout{\string\citation{#2}}\fi
  \@tempcnta\z@\@tempcntb\m@ne\def\@citea{}\@cite{\@for\@citeb:=#2\do
    {\@ifundefined
       {b@\@citeb}{\@citeo\@tempcntb\m@ne\@citea\def\@citea{,}{\bf ?}\@warning
       {Citation `\@citeb' on page \thepage \space undefined}}%
    {\setbox\z@\hbox{\global\@tempcntc0\csname b@\@citeb\endcsname\relax}%
     \ifnum\@tempcntc=\z@ \@citeo\@tempcntb\m@ne
       \@citea\def\@citea{,}\hbox{\csname b@\@citeb\endcsname}%
     \else
      \advance\@tempcntb\@ne
      \ifnum\@tempcntb=\@tempcntc
      \else\advance\@tempcntb\m@ne\@citeo
      \@tempcnta\@tempcntc\@tempcntb\@tempcntc\fi\fi}}\@citeo}{#1}}
\def\@citeo{\ifnum\@tempcnta>\@tempcntb\else\@citea\def\@citea{,}%
  \ifnum\@tempcnta=\@tempcntb\the\@tempcnta\else
  {\advance\@tempcnta\@ne\ifnum\@tempcnta=\@tempcntb \else\def\@citea{--}\fi
    \advance\@tempcnta\m@ne\the\@tempcnta\@citea\the\@tempcntb}\fi\fi}
\makeatother

\newcommand\pubnumber{SLAC--PUB--16103}
\newcommand\pubdate{\today}

\newcommand\pubblock{\rightline{\begin{tabular}{l} \pubnumber\\
         \pubdate \end{tabular}}}
\newenvironment{Abstract}{\begin{quotation} \begin{center}
                       ABSTRACT
     \end{center}\bigskip  }{\end{quotation}}

\def\Acknowledgements{\bigskip  \bigskip \begin{center} \begin{large}
             \bf ACKNOWLEDGEMENTS \end{large}\end{center}}

\def\max{{\mbox{\scriptsize max}}}
\def\ben{\begin{enumerate}}
\def\een{\end{enumerate}}

\begin{document}

\begin{titlepage}
\title{Perturbative Unitarity Constraints on the\\NMSSM Higgs Sector}
\author{Kassahun Betre$^a$, Sonia El Hedri$^{a,b}$ and Devin~G.~E.~Walker$^a$}
\address{$^a$SLAC National Accelerator Laboratory, 2575 Sand Hill Road, Menlo Park, CA 94025, U.S.A.}
\address{$^b$Institut f\"ur Physik (THEP) Johannes Gutenberg-Universit\"at,
D-55099, Mainz, Germany}
\pubblock

\maketitle

\begin{Abstract}
We place perturbative unitarity constraints on both the dimensionful and dimensionless parameters in the Next-to-Minimal Supersymmetric Standard Model (NMSSM) Higgs Sector. These constraints, plus the requirement that the singlino and/or Higgsino constitutes at least part of the observed dark matter relic abundance, generate upper bounds on the Higgs, neutralino and chargino mass spectrum. Requiring higher-order corrections to be no more than 41\% of the tree-level value, we obtain an upper bound of 20~TeV for the heavy Higgses and 12~TeV for the charginos and neutralinos outside defined fine-tuned regions.   If the corrections are no more than 20\% of the tree-level value, the bounds are  7~TeV for the heavy Higgses and 5~TeV for the charginos and neutralinos.  
In all, by using the NMSSM as a template, we describe a method which replaces naturalness arguments with more rigorous perturbative unitarity arguments to get a better understanding of when new physics will appear.

\end{Abstract}
\vfill
\vfill
\end{titlepage}
\def\thefootnote{\fnsymbol{footnote}}
\setcounter{footnote}{0}
\tableofcontents
\newpage
\setcounter{page}{1}

\section{Introduction}
The discovery of the Higgs boson~\cite{Aad:2012tfa,Chatrchyan:2012ufa} completes the experimentally successful Standard Model (SM).  The SM however cannot account for the observed dark matter (DM) in the universe.  Numerous astrophysical observations require 
DM to be electrically neutral, colorless, non-relativistic at redshifts of $z \sim 3000$ and generate the following relic abundance~\cite{Bertone:2004pz,Ade:2013zuv},
\begin{equation}
\Omega_\chi h^2 = 0.1199 \pm 0.0027. \label{eq:relic}
\end{equation}
DM that was once in thermal contact with the SM is very well motivated~\cite{Bertone:2004pz}.  For thermal DM, the observed relic abundance is correlated with the annihilation cross section,
\begin{align}
\Omega_\chi h^{2}\,\simeq \frac{0.1 \,\,\mathrm{pb} \cdot c}{\langle \sigma v \rangle} && \langle \sigma v \rangle \simeq \frac{g^4}{ 8\pi } \frac{1}{ m_\mathrm{\chi}^2}.
\end{align}
Griest and Kamionkowski~\cite{Griest:1989wd} applied unitarity arguments to the annihilation cross section in order to place an 
upper bound of 120 TeV on the dark matter mass~\cite{Profumo:2013yn}.  In this work, we 
show how perturbative unitarity arguments can be used to place bounds, not only on the DM mass, but also on the new particles (``mediators") 
that are associated with the dark matter annihilation and often have SM quantum numbers.  Because of the strength of the thermal dark matter paradigm, we argue the constraints on the mediators offer the strongest estimate now available for when new physics that couples significantly to the SM will appear~\cite{Walker:2013hka}.  In addition, we want to use our methodology to place stronger bounds on the dark matter mass.  In this work, we use the Next-to-Minimum Supersymmetric Standard Model (NMSSM)~\cite{Fayet:1974pd,Drees:1988fc,Betre:2014sra} as a template to implement our philosophy.
%

Throughout the development of the SM, perturbative unitarity arguments have reliably answered the question of when is new physics going to appear.  %
For example in Fermi effective theory, perturbative unitarity is violated around 350 GeV.  New physics in the form of the $W$ and $Z$ bosons (at 80 and 90 GeV, respectively) rescued the theory from becoming strongly coupled.  Moreover without the 
SM Higgs boson, the SM violates perturbative unitarity around %
1.2 TeV~\cite{Dicus:1992vj,Lee:1977eg}.  
The Higgs boson prevented $WW$ scattering from becoming strongly coupled.  Applying perturbative unitarity arguments to models of new physics 
is often not as straightforward as these SM cases.  In most models of new physics beyond the Standard Model, there are many new particles, masses and couplings.  
In order to constrain models of new physics, we employ several low-energy observables as constraints.  In this work, we show how the relic abundance constraint as well as the requirement of having the measured Higgs mass can be used in %
concert with perturbative unitarity arguments to set bounds.  We note there is a deeper connection between perturbative unitarity and a consistent theory of thermal dark matter.  
When the dark matter mass is large and near the perturbativity bounds, the couplings in the annihilation cross section are by definition large.  Thus, 
while annihilating to set the relic abundance while falling out of thermal equilibrium, the dark matter is often confining to form bound states.  %
%
%
These bound states transform trivially under the dark matter stabilization symmetry and are rarely stable over the lifetime of the universe.  Thus, dark matter theories that are near or violate the perturbative unitarity bounds are often inconsistent with observation of the dark matter relic abundance.
 
%
%
The NMSSM 
provides a compelling framework to implement our philosophy.  It naturally explains the observed SM Higgs mass~\cite{Aad:2012tfa,Chatrchyan:2012ufa}, provides a viable dark matter candidate, features gauge coupling unification and solves the hierarchy problem~\cite{Ellwanger:2009dp,Maniatis:2009re}.  We focus on the NMSSM Higgs sector.  To date, none of the new particles predicted by the NMSSM or any new physics model have been found.  Within the particle physics community, this has cause a reconsideration of naturalness, the dominant explanation of when new physics will be bound.  In this work, we aim to provide an alternative to naturalness as a mean to estimate new particle masses.  

In the following, we first provides an overview of the NMSSM Higgs sector. Section~3 details our perturbative unitarity constraints on NMSSM Higgs sector.  
Section~4 discusses our relic abundance constraints. Our results are given in Section~5. We conclude in Section~6.%

\section{The NMSSM}
\label{Sec: NMSSM}
This section provides a brief overview of the NMSSM, using the notations introduced in~\cite{Maniatis:2009re}. The NMSSM superpotential is given by:
	\beq
		W_{NMSSM} = y_u\tilde{u}^*_R(\tilde{Q}^T\epsilon H_u)-y_d\tilde{d}^*_R(\tilde{Q}^T\epsilon H_d)-y_e\tilde{e}^*_R(\tilde{L}^T\epsilon H_d) + \lambda S(H_u^T\epsilon H_d) + {1 \over 3}\kappa S^3.
	\eeq{Eq:nmssmpotential}
Here $\epsilon$ is the $SU(2)$ invariant antisymmetric tensor with the non-zero components given by $\epsilon_{12} = 1, \epsilon_{21} = -1$. We will restrict our discussion to the scalar Higgs, charged higgsino, and neutralino sectors. The winos, binos and squarks will be treated as decoupled. We will work with tree-level expressions. Our analysis will not be significantly altered by including loop corrections.

\subsection{Scalar Higgs Sector}
The scalar part of the Higgs potential has three contributions: the $F$-term ($V_F$), $D$-term ($V_D$), and soft SUSY breaking terms ($V_{soft}$).  We assume no CP-violation in the Higgs sector so all the couplings are real.
	\beqa
		V&=&V_F+V_D+V_{soft}\\
		V_F&=&|\lambda|^2|S|^2\left( H_u^{\dagger}H_u+H_d^{\dagger}H_d\right)+|\lambda(H_u^T\epsilon H_d)+\kappa S^2|^2,\nonumber\\
		V_D&=&{1\over 2}g_2^2|H_u^{\dagger}H_d|^2+{1\over8}(g_1^2+g_2^2)\left(H_u^{\dagger}H_u-H_d^{\dagger}H_d \right)^2, \nonumber\\
		V_{soft}&=&m_{H_u}^2H_u^{\dagger}H_u+m_{H_d}^2H_d^{\dagger}H_d+m_S^2|S|^2+\left( \lambda A_{\lambda}(H_U^T\epsilon H_d)S+{1\over3}\kappa A_{\kappa}S^3+c.c \right).\nonumber
	\eeqa{Eq:higgspotential}
Here $g_1$ and $g_2$ are the $U(1)_Y$ and $SU(2)$ coupling constants respectively.
The fields can be parameterized as follows:
	\beq
		H_d=\begin{pmatrix}
                    \dfrac{1}{\sqrt{2}}\left(v_d+h_d+ia_d\right)	\\
                    \medskip
				H_d^-
			\end{pmatrix}
	, \quad H_u = \begin{pmatrix}
				H_u^+\\
                                \medskip
                                \dfrac{1}{\sqrt{2}}\left(v_u+h_u+ia_u\right)
			\end{pmatrix}
		,
	\eeqn
	\beq
        S = \dfrac{1}{\sqrt{2}}\left(v_s+h_s+ia_s\right).
	\eeqn
        The CP-even fields $h_d, h_u, h_s$ are the scalar Higgses, with vacuum expectation values (VEV) $v_u$, $v_d$ and $v_s$ respectively. The Higgs sector also includes three CP-odd fields $a_u$, $a_d$ and $a_s$  and two charged fields $H^+_u$ and $H^-_d$. Before electroweak symmetry breaking (EWSB), the free parameters of the Higgs NMSSM potential are
	\beq
		\lambda, \; \kappa, \; A_{\lambda}, \; A_{\kappa}, \; m_{H_d}^2, \; m_{H_u}^2, \; m_S^2.
	\eeq{Eq:fullparams}
EWSB requires the Higgs potential to have a global minimum when the Higgs fields $H_u$, $H_d$ and $S$ are at their respective VEVs $v_u$, $v_d$ and $v_s$. 
The vacuum expectation values of the fields must sit at the minimum of the Higgs potential for a successful electroweak symmetry breaking. Evaluated at the vevs, the partial derivatives of the potential with respect to each of the six fields must be zero giving rise to what are called the {\it tadpole} conditions. The stationary condition of the potential with respect to the pseudoscalars gives the following conditions:
	\beqa
		m_{H_d}^2&=& -\mu^2-2{\lambda^2\over g^2}m_Z^2\sin^2(\beta)-{1\over2}m_Z^2\cos(2\beta)+\mu\left( {\kappa\over\lambda}\mu+A_{\lambda}\right)\tan(\beta),\nonumber\\
		m_{H_u}^2&=&-\mu^2-2{\lambda^2\over g^2}m_Z^2\cos^2(\beta)+{1\over2}m_Z^2\cos(2\beta)+\mu\left( {\kappa\over\lambda}\mu+A_{\lambda}\right)\cot(\beta),\nonumber\\
		m_S^2&=&{\lambda^2\over g^2}m_Z^2\left( -2+2{\kappa\over\lambda}\sin(2\beta)+{A_{\lambda}\over\mu}\sin(2\beta)\right)-{\kappa\over\lambda}\mu\left( {\kappa\over\lambda}\mu+A_{\kappa}\right),\nonumber
	\eeqa{Eq:softmasses}	
where $\mu = \lambda v_s/\sqrt{2}$.
These VEVs have to be such that
\begin{align}
    v_u^2 + v_d^2 = v^2 
\end{align}
where $v$ is the SM Higgs VEV, defined as
\begin{align}
    v = 246\,\mathrm{GeV}
\end{align}
Defining an angle $\beta$ such that 
\begin{align}
    v_u = v\sin\beta \quad\quad v_d = v \cos\beta
\end{align}
the EWSB constraints then bring the number of free parameters down to six.
	\beq
		\lambda, \quad \kappa, \quad A_{\lambda}, \quad A_{\kappa}, \quad \tan\beta, \quad \mu.
	\eeq{Eq:param6}
After EWSB, the $W^\pm$ and $Z$ gauge bosons acquire longitudinally polarized components by eating the Goldstone bosons of the up and down Higgses, $G^0$ and $G^\pm$.
The NMSSM scalar Higgs sector is then composed of three scalar CP-event Higgses $h_u$, $h_d$ and $s$, two CP-odd pseudoscalar Higgses $a$ and $a_s$, and one charged Higgs $H^\pm$. The fields $a$ and $H^\pm$ are such that
	\beq
		\left(
			\begin{array}{c}
				a_d\\a_u\\a_s
			\end{array}
		\right) = 
		\left(
			\begin{array}{ccc}
				\cos\beta&\sin\beta&0\\
				-\sin\beta&\cos\beta&0\\
				0&0&1
			\end{array}
		\right)
		\left(
			\begin{array}{c}
				G^0\\a\\a_s
			\end{array}
		\right)
	\eeqn
        and 
	\beq
		\left(
			\begin{array}{c}
				(H_d^-)^*\\H_u^{+}
			\end{array}
		\right)
		=
		\left(
			\begin{array}{cc}
				\cos\beta&\sin\beta\\
				-\sin\beta&\cos\beta
			\end{array}
		\right)
		\left(
			\begin{array}{c}
				G^+\\
				H^+
			\end{array}
		\right).
	\eeqn

        However, at very high energies, the Goldstone Boson Equivalence Theorem (GBET) tells us that the couplings of $W^\pm$ and $Z$ are equal to the couplings of the corresponding Goldstone bosons. In our study, at high center of mass energies, it will then be sufficient to work with $a_d$, $a_u$ ($H^\pm_u$, $(H_d^\mp)^*$), or $G^0$ ($G^\pm$) in place of $Z^0$ ($W^\pm$).

\subsection{Scalar Masses}
The Higgs potential shown in~\leqn{Eq:higgspotential} includes mixing terms for both the scalars and the pseudoscalars. The pseudoscalar mass matrix in the $(a, a_s)$ basis is
	\beq
		M_a^2=
			\left(
				\begin{array}{cc}
					{2\mu\over\sin2\beta}\left( A_{\lambda} + {\kappa\over\lambda}\mu\right)&{\sqrt{2}\lambda\over g}m_Z\left( A_{\lambda}-{2\kappa\over\lambda}\mu\right)\\
					{\sqrt{2}\lambda\over g}m_Z\left( A_{\lambda}-{2\kappa\over\lambda}\mu\right)&{\lambda^2\over g^2}m_Z^2\left( {A_{\lambda}\over\mu}+{4\kappa\over\lambda}\right)\sin2\beta-{3\kappa\over\lambda}A_{\kappa}\mu
				\end{array}
			\right)
	\eeq{oddmass}	
In the $(h_d, h_u, h_s)$ basis, the matrix elements of the scalar mass matrix are
	\beqa
		M^2_{h,11}&=&m_Z^2\cos^2\beta+\mu\left( {\kappa\over\lambda}+A_{\lambda}\right)\tan\beta,\\
		M^2_{h,22}&=&m_Z^2\sin^2\beta+\mu\left( {\kappa\over\lambda}+A_{\lambda}\right)\cot\beta,\nonumber\\
		M^2_{h,33}&=&{4\kappa^2\over\lambda^2}\mu^2+{\kappa\over\lambda}A_{\kappa}\mu+{\lambda^2\over g^2}{A_{\lambda}m_Z^2\over\mu}\sin2\beta,\nonumber\\
		M^2_{h,12}&=&2\left( {\lambda^2\over g^2}-{1\over4}\right)m_Z^2\sin2\beta-\mu\left({\kappa\over\lambda}\mu A_{\lambda} \right),\nonumber\\
		M^2_{h,13}&=&{2\sqrt{2}\lambda\over g}\mu m_Z \cos\beta - {\sqrt{2}\lambda m_Z\over g}\left( A_{\lambda}+{2\kappa\over\lambda}\mu\right)\sin\beta,\nonumber\\
		M^2_{h,23}&=&{2\sqrt{2}\lambda\over g}\mu m_Z \sin\beta - {\sqrt{2}\lambda m_Z\over g}\left( A_{\lambda}+{2\kappa\over\lambda}\mu\right)\cos\beta.\nonumber
	\eeqa{evenmass}

        The NMSSM Higgs sector then includes three scalar mass eigenstates $h_1$, $h_2$, $h_3$, two pseudoscalar mass eigenstates $a_1$ and $a_2$ and one charged mass eigenstate $H^\pm$. 
        
        In this paper, we require the lightest CP-even Higgs mass eigenstate to be the $125\,\mathrm{GeV}$ Higgs boson mass measured at the LHC in 2012~\cite{Aad:2012tfa,Chatrchyan:2012ufa}. This additional constraints allows us to fix one of the parameters shown in~\leqn{Eq:param6} and brings the number of free parameters down to five. This is achieved by solving the characteristic equation, $\text{\bf Det}[M^2_h - m_h^2{\bf I}] = 0$ for $A_{\kappa}$. Now there are only 5 independent parameters left,
	\beq
		\lambda, \quad \kappa, \quad A_{\lambda}, \quad \tan\beta, \quad \mu.
	\eeq{Eq:param5}
        
Since we operate at tree-level, the NMSSM Higgs sector parameters further obeys the constraint
        \begin{align}
            m_{h_1}^2 \lesssim m_Z^2\, \left( \cos^2 (2\beta)  + {2 \bigl| \lambda \bigr|^2 \sin^2 (2\beta) \over g_1^2 + g_2^2} \right),
        \end{align}
        which leads to a lower bound on $\lambda$
        \begin{align}
            \lambda \gsim 0.8.
        \end{align}

Rotation by $\beta$ also rotates the charged Higgses, $((H_d^-)^*,H_u^+)$ into the Goldstone boson $w^+$ that gets eaten by the $W^+$ boson and the physical charged Higgs $H^+$.
	\beq
		\left(
			\begin{array}{c}
				(H_d^-)^*\\H_u^{+}
			\end{array}
		\right)
		=
		\left(
			\begin{array}{cc}
				\cos\beta&\sin\beta\\
				-\sin\beta&\cos\beta
			\end{array}
		\right)
		\left(
			\begin{array}{c}
				w^+\\
				H^+
			\end{array}
		\right).
	\eeqn
The mass of the charged Higgs mass eigenstate given by
	\beq
		m_{H^\pm}^2=m_W^2-{2\lambda^2\over g^2}m_Z^2+{2\mu\over\sin2\beta}\left( A_{\lambda}+{\kappa\over\lambda}\mu\right).
	\eeq{chargemass}
The charged particles, $w^-$ and $H^-$ are the conjugates of the above.

\subsection{Neutralino Sector}
The neutral $SU(2)\times U(1)$ gauginos $\tilde{B}$ and $\tilde{W}^3$ generally mix with the higgsinos $\tilde{H}_u$, $\tilde{H}_d$ and the singlino $\tilde{S}$. The $5\times5$ neutralino mass matrix in the basis $(\tilde{B}^0,\tilde{W}^3,\tilde{H}_d^0,\tilde{H}_u^0,\tilde{S})$, is then
	\beq
		M_{\tilde{\chi}^0}=
			\begin{pmatrix}
				M_1&0&-c_{\beta}s_Wm_Z&s_{\beta}s_Wm_Z & 0\\
				0&M_2&c_{\beta}c_Wm_Z&-s_{\beta}c_Wm_Z&0\\
				-c_{\beta}s_Wm_Z&c_{\beta}c_Wm_Z&0&-\mu&-\lambda v_u\\
				s_{\beta}s_Wm_Z&-s_{\beta}c_Wm_Z&-\mu&0&-\lambda v_d\\
                                0&0&-\lambda v_u&-\lambda v_d&\sqrt{2}\kappa v_s
			\end{pmatrix}
	\eeqn 
Since we decouple the winos and binos, we need only to consider the $3\times 3$ Higgsino/singlino block of this mixing matrix
	\beq
		M_{\tilde{\chi}^0}=
			\begin{pmatrix}
				0&-\mu&-\lambda v_u\\
				-\mu&0&-\lambda v_d\\
                                -\lambda v_u & -\lambda v_d &\sqrt{2}\kappa v_s
			\end{pmatrix}
	\eeqn 
        Besides the neutralinos, the NMSSM also includes two charginos $\tilde{C}_{12}^\pm$ whose squared masses are
	\begin{align}
            m^2_{\tilde{C}_1},m^2_{\tilde{C}_2} &= {1\over2}\left(\vphantom{\sqrt{(M_2)^2}} M_2^2+\mu^2+2m_W^2\right.\\
                &\left.\mp\sqrt{(M_2^2+\mu^2+2m_W^2)^2-4(\mu M_2-m_W^2\sin2\beta)^2}\right)
	\end{align}
        Decoupling the winos only leaves one chargino, the charged Higgsino $\tilde{H}^\pm$, whose mass is 
        \begin{align}
            m_{\tilde{H}^\pm} &= \mu.
        \end{align}
	\subsection{The heavy limit}
    Cases of particular interest in our study are configurations where 
    \begin{align}
        \mu, A_\lambda, A_\kappa \gg v.
    \end{align}
    In these cases, the scalar and fermion mass spectra simplify considerably. In particular, the mixing between the Higgs/Higgsino and singlet/singlino sectors becomes negligible. The masses of the non-decoupled scalar particles then become
    \begin{align}
        m^2_{h_1} &= 125\,\mathrm{GeV}\\
        m^2_{h_2} &= m^2_{a_1} = m^2_{H^\pm} = \frac{2\mu}{\sin(2\beta)} \left(\frac{\kappa}{\lambda}\mu + A_\lambda\right)\\
        m^2_{h_3} &= 4\frac{\kappa^2}{\lambda^2}\mu^2 + A_\kappa\frac{\kappa}{\lambda}\mu \quad \quad m^2_{a_2} = -3 \frac{\kappa}{\lambda}\mu A_\kappa
    \end{align}
    while the fermion masses are
    \begin{align}
        m_{\tilde{H}_1^0} &= m_{\tilde{H}_2^0} = m_{\tilde{H}^\pm} = \mu \quad\quad m_{\tilde{S}} = \frac{2\kappa}{\lambda}\mu
    \end{align}
    Since $\lambda \sim \mathcal{O}(1)$, the NMSSM Higgs sector in the heavy limit has the three following characteristic scales
    \begin{align}
        \mu , \; \sqrt{A_\lambda \mu}, \; \sqrt{A_\kappa \mu}.
    \end{align}
    The energy scale $\mu$ has a crucial role since it sets the energy scale of the fermionic sector and is involved in all the scalar and pseudoscalar Higgs masses. Notably, in order for the scalar sector to be much heavier than the fermion sector, we would need
    \begin{align}
        A_\lambda \gg \mu \quad\text{or}\quad A_\kappa \gg \mu.
        \label{Eq: heavyHiggs}
    \end{align}
    We will show later in this paper that these regions of parameter space are tightly constrained by unitarity.

\section{Vacuum Constraints}

The Higgs potential shown in~\leqn{Eq:higgspotential} generally has a large number of minima. Requiring the EWSB vacuum shown in~\leqn{Eq:softmasses} to be stable -- deeper than all the other vacua -- provides preliminary constraints on the 5 parameters in~\leqn{Eq:param5}. 

At the EWSB vacuum, the Higgs potential takes the following value
	\beq
		V_{min} = -\lambda^2{m_Z^4\sin^22\beta\over g^4}-{m_Z^4\cos^22\beta\over 2g^2}+\bar{V}_{min},
	\eeq{vmin}
where,
	\beq
		\bar{V}_{min}={\kappa^2\over\lambda^4}\mu^4+{2\over3}{\kappa\over\lambda^3}A_{\kappa}\mu^3+{1\over\lambda^2}m_S^2\mu^2.
	\eeq{vbarmin}
        Although the Higgs potential~\leqn{Eq:higgspotential} cannot be analytically minimized, five different classes of vacua other than the EWSB vacuum can be identified, as shown in~\cite{Kanehata:2011ei}. These are
\begin{itemize}
    \item $|H_u| = |H_d| = |S| = 0$
    \item $|H_u| = |H_d|\neq 0$ and $|S| = 0$
    \item $|H_u| \neq 0$ and $|H_d| = |S| = 0$
    \item $|H_d| \neq 0$ and $|H_u| = |S| = 0$
    \item $|S| \neq 0$ and $|H_u| = |H_d| = 0$.
\end{itemize}
In our study, we locate the minima in these directions and require the EWSB vacuum to be deeper than any of these vacua.
A more complete search for the global minima of the NMSSM Higgs potential using algebraic approaches is done in~\cite{Maniatis:2006jd}. They found that for both CP violating and CP conserving Higgs potential, a large part of the parameter space is eliminated when the EWSB minimum is required to be a global minimum.

\section{Constraints from Perturbative Unitarity}
\label{Sec: Unitarity}
One fundamental constraint on the SM and on models of new physics is the fact that all the scattering amplitudes in the theory must be unitary. Applying this unitarity criterium on the NMSSM Higgs sector would allow to constrain not only the dimensionless couplings $\lambda$ and $\kappa$ but also ratios of energy scales.
\subsection{Overview of unitarity}
\begin{figure}
\vskip -0.4in
\centering
\includegraphics[width=0.8\linewidth]{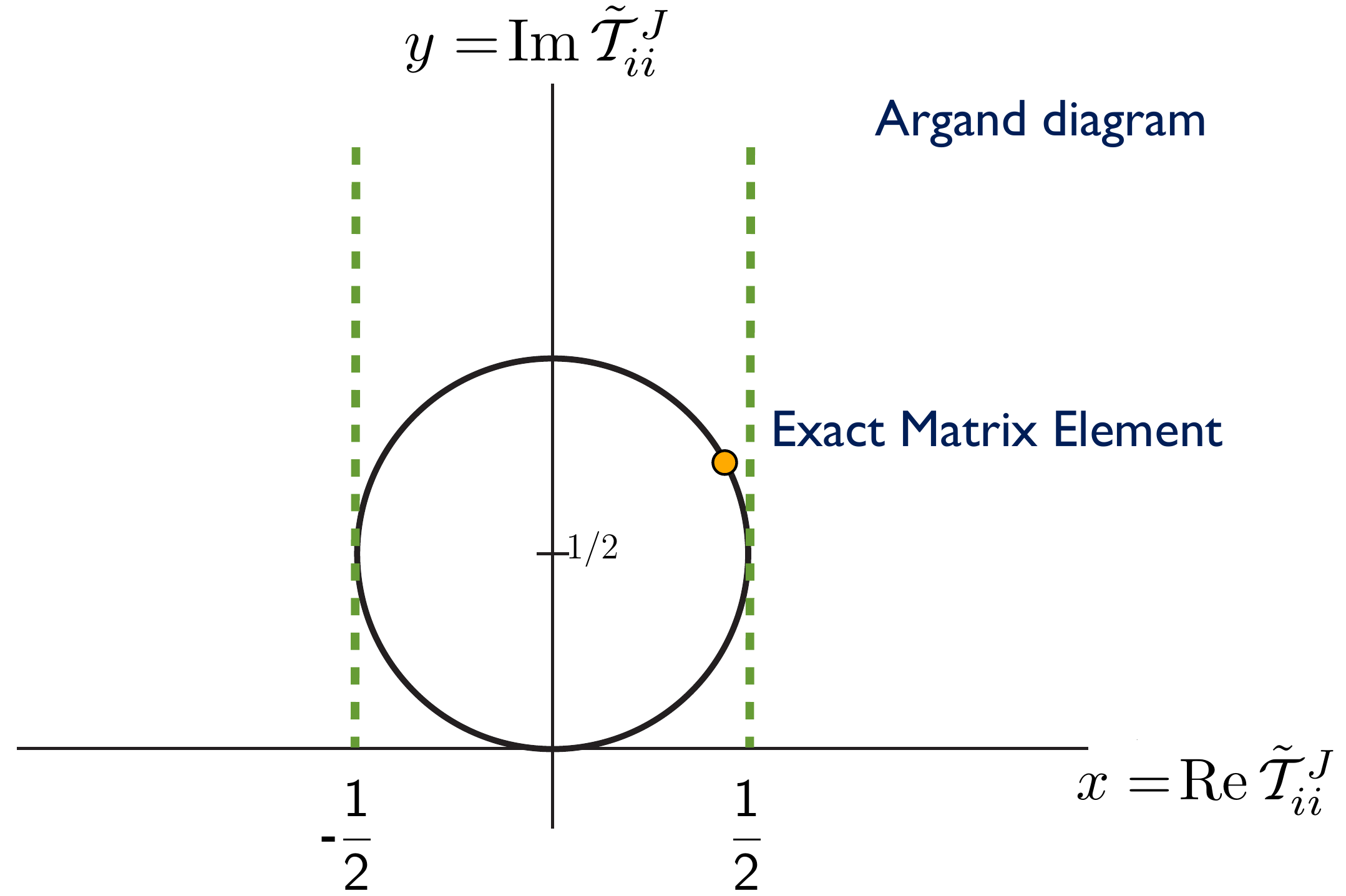}
\caption{The Argand circle. In unitary theories, the eigenvalues of the exact partial-wave components of the scattering matrix, $\mathcal{T}^J_{ii}$, must lie on this circle.}
\label{Fig: argandExact}
\end{figure}

When a scattering takes place, the evolution of the different states from $t = -\infty$ (incoming) to $t = +\infty$ (outgoing) is described by the scattering S-matrix.
This matrix is usuallty decomposed in function of a $T$-matrix such as
\begin{align}
    S = \mathbf{1} + i T.
\end{align}
 In this study, we focus on two-to-two scattering processes, where the scattering amplitudes can be expressed in function of the center of mass energy $\sqrt{s}$ and the scattering angle $\theta$. For these processes and for a given initial state $\langle i |$ and final state $|j \rangle$, the matrix element $\langle f | T | i \rangle$ is related to the scattering amplitude $\mathcal{M}_{fi}$ through
	\beq
        \bra{f}T\ket{i}=(2\pi)^4\delta^4(p_f-p_i)\mathcal{M}_{fi}(\sqrt{s},\cos\theta),
	\eeq{feynmanME}
Requiring that $S$ should be unitary leads to a non-trivial constraint on the $T$-matrix. 
	\begin{align}
		S^{\dagger}S = \One\quad \implies -i(T-T^{\dagger}) =T^{\dagger}T
                \label{Eq: ssunitary}
	\end{align}
        For a given initial state $\langle i |$ and final state $|f\rangle$, the matrix element $\langle f| T|i\rangle$ can be computed using Feynman rules
        \begin{align}
            \langle f| T|i\rangle = (2\pi)^4 \delta^4(p_f - p_i)\mathcal{T}_{fi}.
        \end{align}
        ~\leqn{Eq: ssunitary} then becomes
        \begin{align}
            -i \left(\mathcal{T}_{ji} - \mathcal{T}^*_{ij} = \sum_n \mathcal{T}^*_{nj}\mathcal{T}_{ni}\right).
            \label{Eq: sunitary}
        \end{align}
        Going further requires simultaneously diagonalizing both sides of ~\leqn{Eq: sunitary}. This can be performed by decomposing the $\mathcal{T}_{ij}$ matrix into its angular momentum eigenstates. For two-to-two scattering, these eigenstates are given by
	\beq
        \mathcal{T}_{ij}^J(\sqrt{s})={1\over2}{\lambda_i^{1/4}\lambda_f^{1/4}\over16\pi s}\int_{-1}^1d(\cos\theta)\mathcal{T}_{ij}(\sqrt{s},\cos\theta)P^J(\cos\theta).
	\eeqn
        where the $\lambda(x, y, z) = x^2 + y^2 + z^2 - 2 xy - 2 yz - 2 xz$ are phase space factors. For each of the $\mathcal{T}^J_{ij}$ components, simultaneously diagonalizing both sides of ~\leqn{Eq: sunitary} gives
    \begin{align}
        -i \left(\mathcal{T}_{ii}^J - \mathcal{T}^{*J}_{ii}\right) &= \sum_n \mathcal{T}^{*J}_{ni}\mathcal{T}_{ni}^J
    \end{align}
    which leads to the following requirement on the eigenvalues of $\mathcal{T}^J$
    \begin{align}
        2 \mathrm{Im}\mathcal{T}^J_{ii} &= \left|\mathcal{T}^J_{ii}\right|^2.
        \label{Eq: optic}
    \end{align}
    This identity is known as the optical theorem. Decomposing the right hand side of ~\leqn{Eq: optic} into real and imaginary part leads to the following identity
    \begin{align}
        \left(\mathrm{Re}\mathcal{T}^J_{ii}\right)^2 + \left(\mathrm{Im}\mathcal{T}^J_{ii} - \frac{1}{2}\right)^2 &= \frac{1}{2}.
        \label{Eq: circle}
    \end{align}
    Geometrically, ~\leqn{Eq: circle} means that the eigenvalues of $\mathcal{T}^J$ lie on a circle of center $\left(0, \frac{1}{2}\right)$ and radius $R = \frac{1}{2}$ --the Argand circle-- in the complex plane. This identity is illustrated on Fig.~\ref{Fig: argandExact}. It is important to note that ~\leqn{Eq: circle} applies to the exact scattering matrix elements. In perturbation theory, the $\mathcal{T}^J$ matrices can be computed only up to a finite loop order. These approximated $\mathcal{T}^J$ generally do not lie on the Argand circle.

\subsection{Perturbativity and Unitarity}
\label{SubSec: UnitarityOverview}
In a unitary theory, the eigenvalues of all the scattering matrices must lie on the Argand circle. Since this requirement only applies to the fully resummed scattering amplitudes, it is in general impossible to directly apply it. However, Shuessler and Zeppenfeld \cite{Schuessler:2007av} have given a useful prescription to set unitarity bounds based on tree-level scattering amplitudes with the further assumption that the theory is perturbative. 

Tree-level scattering amplitudes are not subject to the unitarity requirement of ~\leqn{Eq: circle}. In fact, as shown in Fig.~\ref{Fig: loopcorr}, since they are real, they have to lie on the x-axis in the complex plane and never reach the Argand circle. The loop corrections then play a crucial role in unitary theories. As shown in Fig.~\ref{Fig: loopcorr}, they bring the scattering amplitudes closer and closer to the Argand circle, often following a circuitous route. Fig.~\ref{Fig: short} shows the most optimistic case, in which the loop corrections take the shortest possible path to the circle and are therefore minimal. 

\begin{subfigures}
\begin{figure}
\vskip -0.4in
\centering
\includegraphics[width=0.8\linewidth]{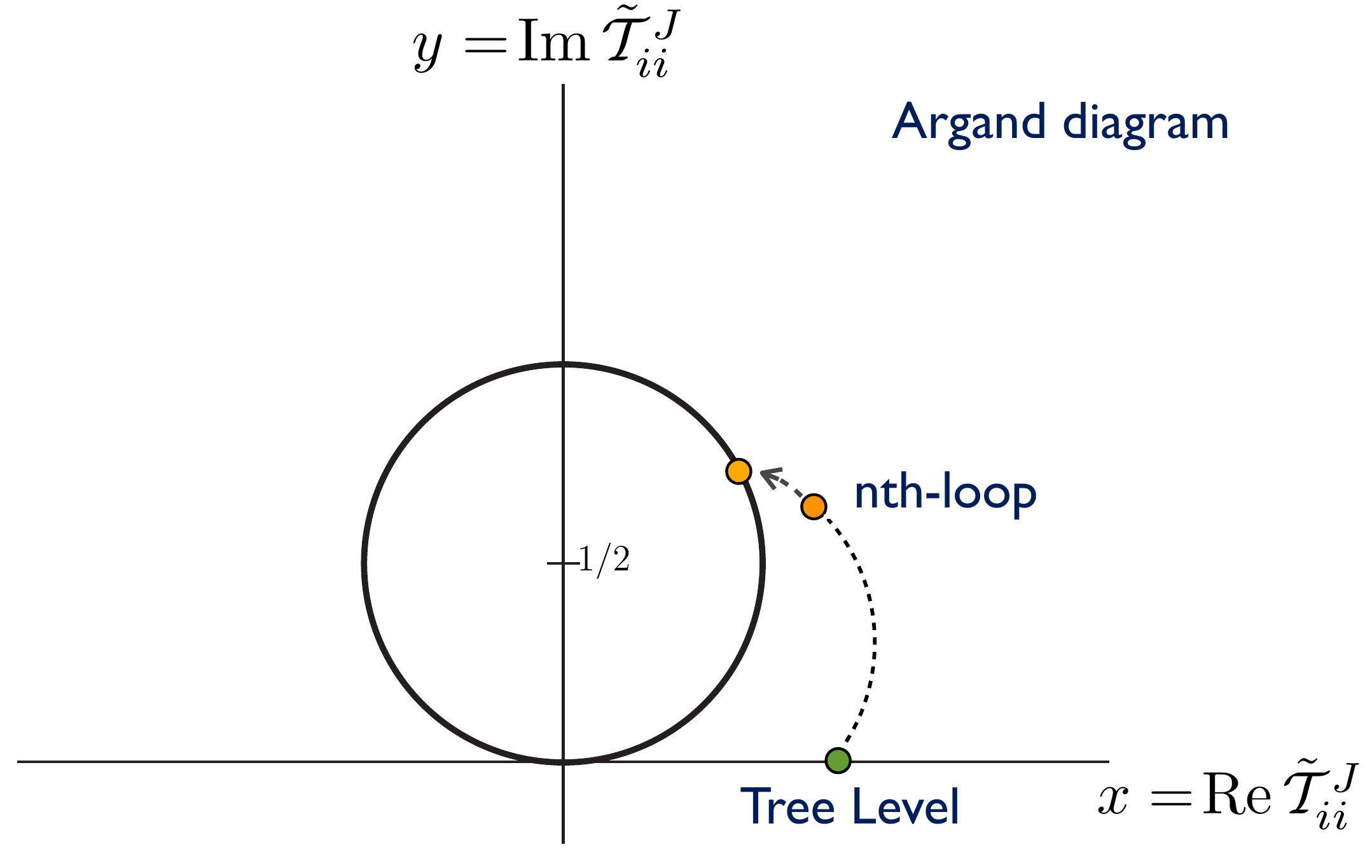}
\caption{Tree-level value of the eigenvalues of $\mathcal{T}^J$. The arrows show how the loop corrections bring the eigenvalues of $\mathcal{T}^J$ closer to the Argand circle.}
\label{Fig: loopcorr}
\end{figure}
\begin{figure}
\vskip -0.4in
\centering
\includegraphics[width=0.8\linewidth]{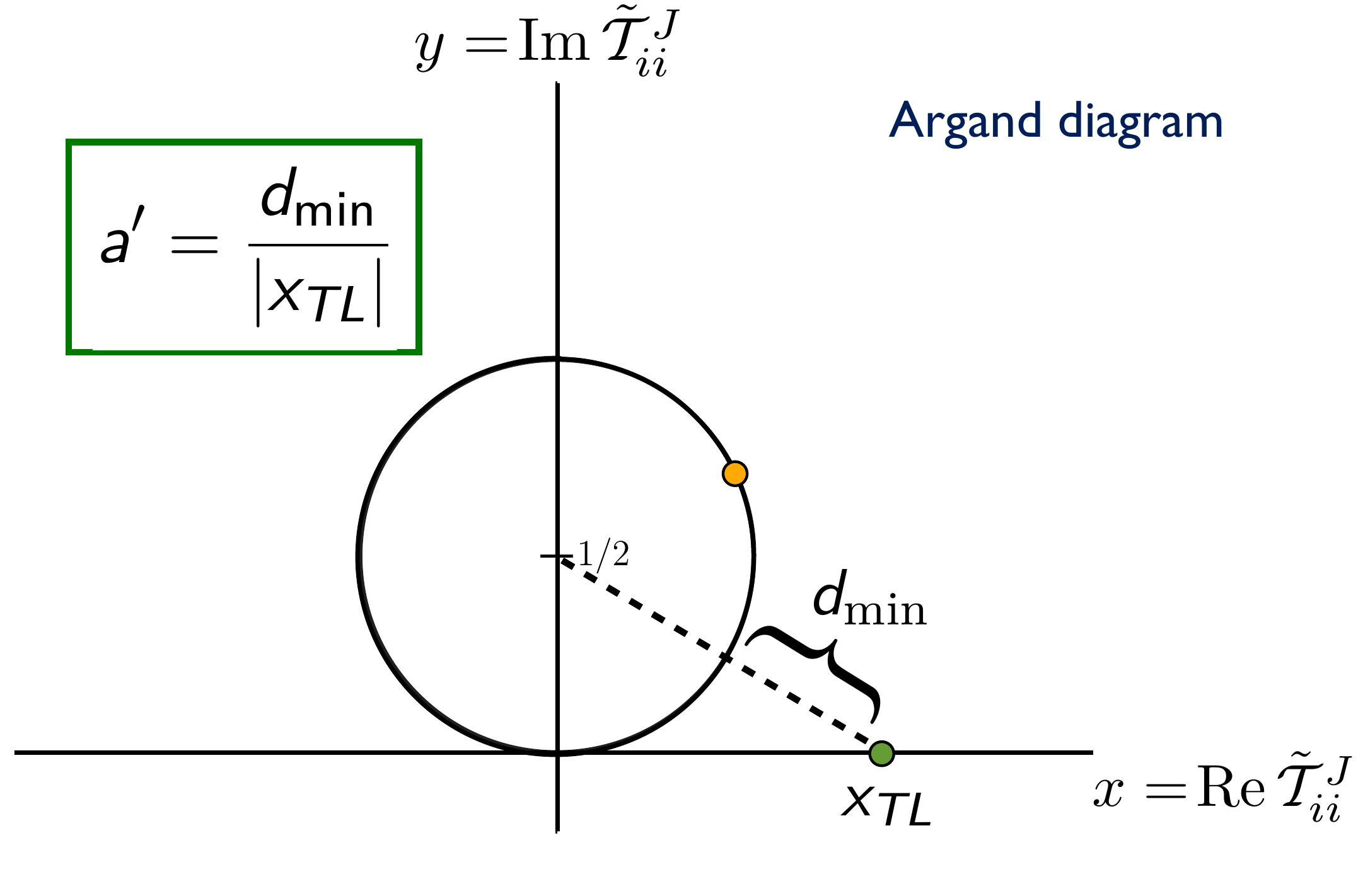}
\caption{Optimistic case where the loop corrections take the shortest route to the Argand circle. The orange point is the tree-level value of the $\mathcal{T}^J$ eigenvalue considered.}
\label{Fig: short}
\end{figure}
\end{subfigures}

In the optimistic case shown in Fig.~\ref{Fig: short}, the size of the loop corrections can be straightforwardly computed using simple geometric arguments. The relative amount of loop corrections with respect to the tree-level value is given by
\begin{align}
    a = \left|\frac{\mathcal{T}^{J, \mathrm{exact}}_{ii} - \mathcal{T}^{J, \mathrm{tree}}_{ii}}{\mathcal{T}^{J, \mathrm{tree}}_{ii}}\right| = \frac{1}{\left|\mathcal{T}^{J, \mathrm{tree}}_{ii}\right|}\left[\sqrt{\left(\mathcal{T}^{J, \mathrm{tree}}_{ii}\right)^2 + \frac{1}{4}} - \frac{1}{2}\right]
    \label{Eq: aval}
\end{align}
The ratio $a$ represents the minimal relative amount of loop corrections needed to unitarize a given theory. Since it assumes that the loop corrections take the most direct route to the circle, this estimate is conservative. 

If $a$ is close to one and the theory is not perturbative. Computing scattering amplitudes at tree-level thus allows to estimate when perturbativity is broken in a unitary theory. 

Setting a maximal $a$, beyond which perturbativity is broken, introduces some amount of arbitrariness in our approach. This arbitrariness is however limited. The following two requirements
\begin{align}
    a \leq 41\% \quad \text{and}\quad a \leq 20\%
\end{align}
correspond respectively to requiring
\begin{align}
    \left|\mathcal{T}^{J, \mathrm{tree}}_{ii}(s)\right|  \leq \frac{1}{2} \quad \text{and}\quad \left|\mathcal{T}^{J, \mathrm{tree}}_{ii}(s)\right|  \leq \frac{1}{4}.
    \label{Eq: limit}
\end{align}
These requirements hold for all center of mass energies $\sqrt{s}$. We show results for these two cases.

Although our study focuses on the NMSSM, the approach outlined here is universal and only assumes that the theory considered is unitary and perturbative. Once a maximal ratio $a_{\mathrm{max}}$ is chosen, upper bounds on the tree-level scattering amplitudes can be derived from~\leqn{Eq: aval} for any type of model.

\subsection{Pole handling}
\label{SubSec: poles}
Sec.~\ref{SubSec: UnitarityOverview} outlined a conservative and universal approach to derive perturbativity requirements in unitary theories. These requirements can be enforced by setting upper bounds on the eigenvalues of the tree-level scattering matrices, as shown in ~\leqn{Eq: limit}. Since these bounds hold for all $\sqrt{s}$, the approach that would lead to the strongest constraints would be to scan over $\sqrt{s}$ and apply ~\leqn{Eq: limit} to the maximal $\left|\mathcal{T}^{J, \mathrm{tree}}_{ii}\right|$. For a given initial state $\langle i |$ and final state $| j\rangle$, a general tree-level scattering matrix element looks like
\begin{align}
    \mathcal{T}_{ij} &= \mathcal{A}^{\mathrm{4-point}} + \sum_n  \frac{\mathcal{A}^s_n}{s - m_n^2} + \sum_n \frac{\mathcal{A}^t_n}{t - m_n^2} + \sum_n \frac{\mathcal{A}^u_n}{u - m_n^2}
\end{align}
where the $\mathcal{A}$ are constants  and the $m_n$ are the masses of the different propagators. Far from $s, t$ or $u$-channel poles, the amplitudes are well-behaved and ~\leqn{Eq: limit} can be straightforwardly applied. In the regions near the poles, however, imaginary contributions to the scattering amplitudes need to be taken into account. These imaginary contributions in general prevent both sides of ~\leqn{Eq: sunitary} from being simultaneously diagonalizible and the approach shown in Sec.~\ref{SubSec: UnitarityOverview} is no longer valid. When scanning over $\sqrt{s}$, regions near the poles then need to be treated with care. Our approach for handling the poles is outlined in the appendix. 

\subsection{Unitarity in the NMSSM Higgs sector}
We now apply the procedure outlined in the previous sections to the NMSSM Higgs sector, focusing on two-to-two scalar scattering. The states we consider in this study are neutral pairs of CP-even scalars. We consider only the $J = 0$ partial-wave components of the scattering matrices, which are expected to give the strongest unitarity bounds.

\subsubsection{$\sqrt{s} \to \infty$ limit}
\begin{figure}
\begin{center}
\vskip -0.4in
\includegraphics[height=4.0in]{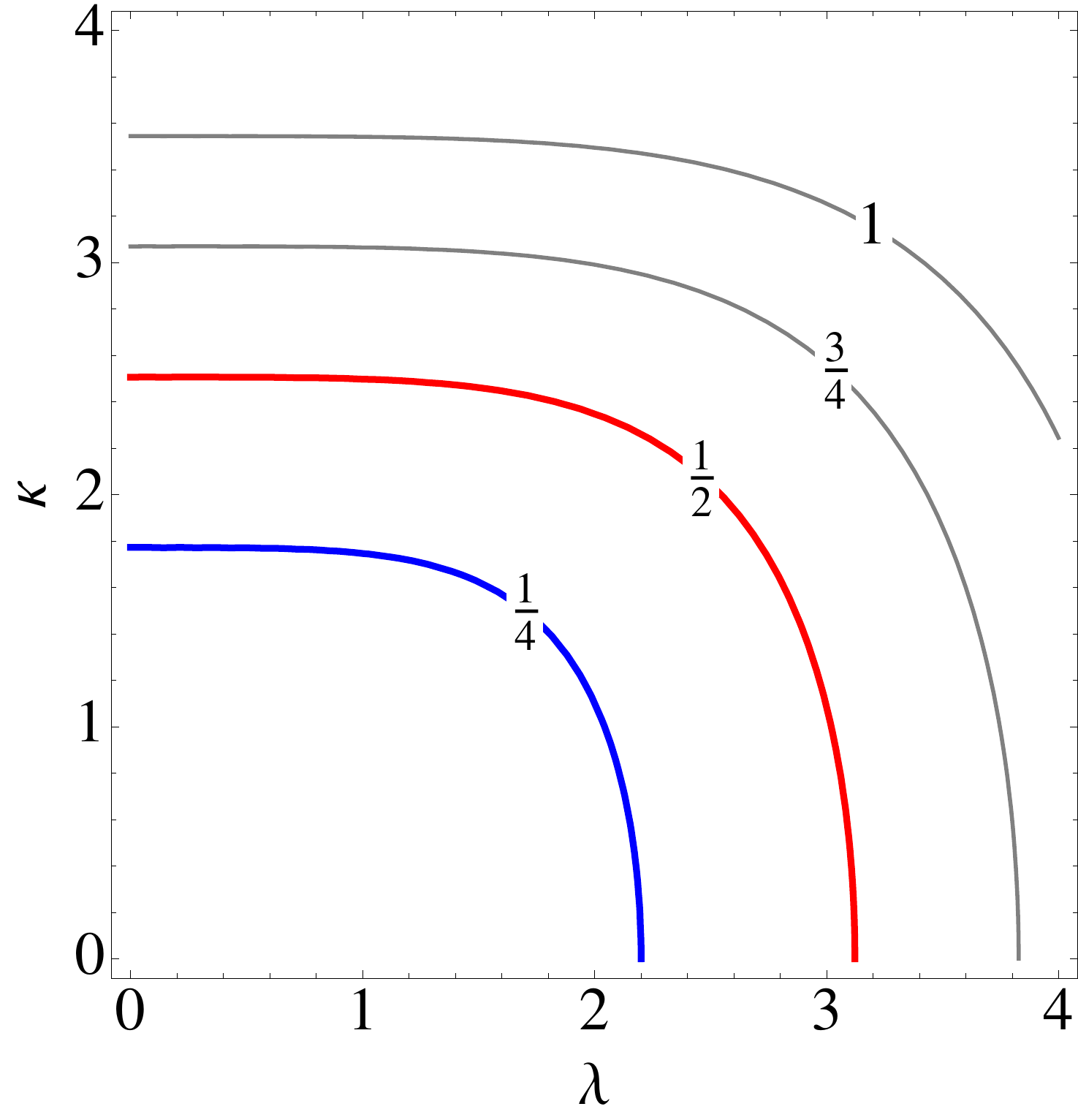}
\caption{Isocontours of the largest eigenvalue of the s-wave scattering matrix $\mathcal{T}^0$ in function of $\lambda$ and $\kappa$ for $\sqrt{s}\to\infty$. The blue and red contours are the upper bounds on $\lambda$ and $\kappa$ for $a_{\mathrm{max}} = 20$ and $41\%$ respectively.}
\label{Fig: dimless}
\end{center}
\end{figure}

A simple preliminary study can be performed by computing the scattering amplitudes for $s\rightarrow \infty$. In this case, the $s$, $t$ and $u$-channels vanish and the contributions to the amplitude entirely come from four-point interactions. The parameters that are constrained by unitarity in this limit are therefore the dimensionless couplings $\lambda$ and $\kappa$. A similar procedure has been performed in the SM by \cite{Dicus:1992vj,Lee:1977eg} and has allowed to set an upper bound on the Higgs quartic coupling, and therefore on the SM Higgs mass.

Since the $s\rightarrow \infty$ amplitudes do not involve propagators, we can work in the interaction eigenbasis. We therefore consider the following 15 CP-even scalar pairs.
	\beqa
		\ket{CP+} &=& \{h_dh_d,\, h_dh_u,\, h_dh_s,\, h_uh_u,\, h_uh_s,\, h_sh_s,\,a_da_d,\, a_da_u,\, a_da_s,\, a_ua_u,\, a_ua_s,\, a_sa_s,\CR
		&&H^-_dH_d^{-*},H_u^+H_u^{+*},\,1/\sqrt{2}(H_d^-H_u^++H_d^{-*}H_u^{+*})\}
	\eeqan
        The strongest constraints will come from the $11\times11$ upper block of this CP-even matrix. This block is shown in Appendix \ref{App: sinf}.

        We could further consider $h_ih_j \to \chi\chi$ or $a_ia_j\to\chi\chi$, where $\chi$ is dark matter or other fermionic particle. However, the dominant contribution to these scatterings occurs for $J=1$ partial waves. Since our study focuses on $S$-wave scattering, we do not take fermion scattering into account in our study.

        The eigenvalues of the s-wave scattering matrix $\mathcal{T}^0$ have to be computed numerically. Fig.~\ref{Fig: dimless} shows isocontours of the largest eigenvalue of $\mathcal{T}^0$ in function of $\lambda$ and $\kappa$. For the values of $a_{\mathrm{max}}$ that we use, the bounds on these dimensionless couplings are
        \begin{align}
            \lambda, \kappa &\lsim 3\quad\text{for}\quad a_{\mathrm{max}} = 41\%\\
            \lambda, \kappa &\lsim 2\quad\text{for}\quad a_{\mathrm{max}} = 20\%
        \end{align}

\subsubsection{$\sqrt{s}$ finite}
\label{subsec:sFinite}
\begin{figure}
    \centering
    \begin{fmffile}{fourscalar}
        \begin{fmfgraph*}(115,115)
            \fmfleft{i1,i2}
            \fmfright{o1,o2}
            \fmf{dashes, label=$\phi_{i1}$,label.side=left}{i1,v1}
            \fmf{dashes, label=$\phi_{f1}$}{v1,o2}
            \fmf{dashes, label=$\phi_{i2}$,label.side=right}{i2,v1}
            \fmf{dashes, label=$\phi_{f2}$,label.side=left}{v1,o1}
        \end{fmfgraph*}
    \end{fmffile}
    \quad\quad
    \begin{fmffile}{sscalar}
        \begin{fmfgraph*}(100,140)
            \fmfbottom{i1,i2}
            \fmftop{o1,o2}
            \fmf{dashes, label=$\phi_{i1}$,label.side=right}{i1,v1}
            \fmf{dashes, label=$\phi_{i2}$,label.side=left}{i2,v1}
            \fmf{dashes, label=$\phi_{p}$,tension=0.5}{v1,v2}
            \fmf{dashes, label=$\phi_{f1}$,label.side=left}{v2,o2}
            \fmf{dashes, label=$\phi_{f2}$}{v2,o1}
        \end{fmfgraph*}
    \end{fmffile}
    \\\vspace{1cm}
    \begin{fmffile}{tscalar}
        \begin{fmfgraph*}(140,100)
            \fmfbottom{i1,i2}
            \fmftop{o1,o2}
            \fmf{dashes, label=$\phi_{i1}$,label.side=right}{i1,v1}
            \fmf{dashes, label=$\phi_{i2}$,label.side=left}{i2,v2}
            \fmf{dashes, label=$\phi_{p}$,tension=0.5}{v1,v2}
            \fmf{dashes, label=$\phi_{f1}$,label.side=right}{v1,o1}
            \fmf{dashes, label=$\phi_{f2}$}{v2,o2}
        \end{fmfgraph*}
    \end{fmffile}
    \quad\quad
    \begin{fmffile}{uscalar}
        \begin{fmfgraph*}(140,100)
            \fmfbottom{i1,i2}
            \fmftop{o1,o2}
            \fmf{dashes, label=$\phi_{i1}$,label.side=right}{i1,v1}
            \fmf{dashes, label=$\phi_{i2}$,label.side=left}{i2,v2}
            \fmf{dashes, label=$\phi_{p}$}{v1,v2}
            \fmf{phantom}{v1,o1}
            \fmf{phantom}{v2,o2}
            \fmffreeze
            \fmf{dashes, label=$\phi_{f1}$,label.side=left,tension=1}{v1,o2}
            \fmf{dashes, label=$\phi_{f2}$,tension=1}{v2,o1}
        \end{fmfgraph*}
    \end{fmffile}
\caption{\label{Fig: scalardiagrams} Feynman diagrams contributing to the scalar scattering matrix}
\end{figure}
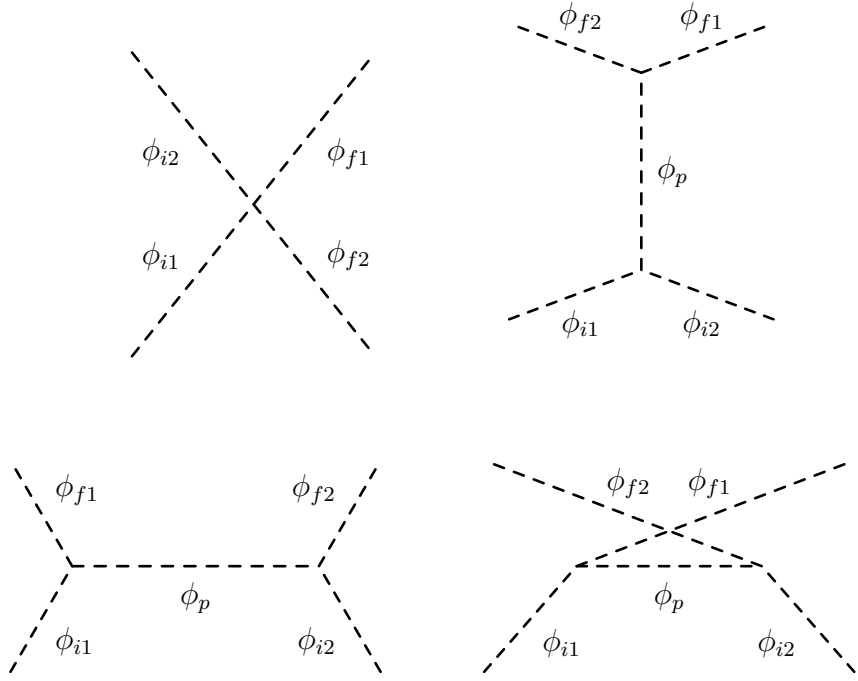

To put unitarity constraints on the trilinear couplings $A_{\lambda}$ and $A_{\kappa}$, we need to evaluate the partial wave scattering amplitudes for finite $\sqrt{s}$. For a process of the form $\phi_{i1}\phi_{i2}\rightarrow\phi_{f1}\phi_{f2}$, the different types of diagrams contributing to the scattering amplitudes are shown in Fig.~\ref{Fig: scalardiagrams}.
Recall that the general form of the total amplitude for two-to-two scattering processes is
\begin{align}
    \mathcal{T}_{ij} &= \mathcal{A}^{\mathrm{4-point}} + \sum_n  \frac{\mathcal{A}^s_n}{s - m_n^2} + \sum_n \frac{\mathcal{A}^t_n}{t - m_n^2} + \sum_n \frac{\mathcal{A}^u_n}{u - m_n^2}
\end{align}
The $s$, $t$ and $u$-channel contributions to the amplitude are then expected to constrain ratios of scales such as $A_\lambda^2/s$ or $A_\kappa^2/s$.

Here, we use the following 15 CP-even pairs of scalar mass eigenstates
\begin{align}
    \big| CP^+\rangle &= \left\{h_1 h_1, h_1 h_2, h_2 h_2, h_1h_3, h_2h_3,h_3h_3, zz, za_1, za_2,a_1a_1, a_1a_2, a_2a_2\right.\\
        &\left., w^+ w^-, H^+H^-, 1/\sqrt{2} \left(H^+w^- + H^-w^+\right) \right\}
\end{align}
Here, the $w^\pm$ and $z$ particles are the Goldstone bosons associated to the longitudinal components of the $W^\pm$ and $Z$ bosons.

\begin{figure}
\begin{center}
\vskip -0.51in
\includegraphics[width=0.8\linewidth]{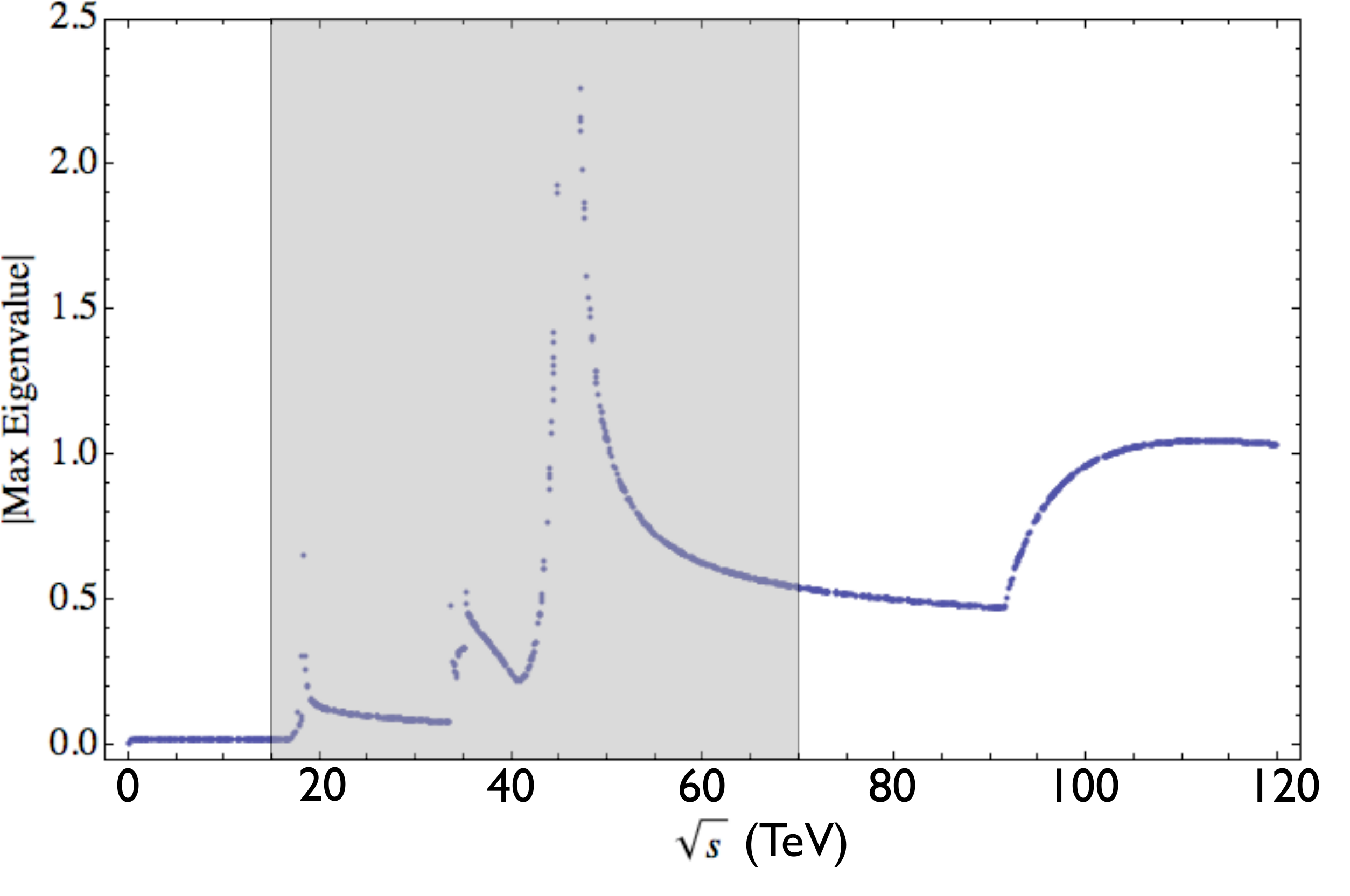}
\caption{Absolute value of the largest eigenvalue as a function of $\sqrt{s} \,\,(\text{TeV})$ for a parameter point with $\lambda = 0.8396, \kappa=2.3410, A_{\lambda}=-6814.50 \,\mathrm{GeV}, A_{\kappa} = -4364.70 \,\mathrm{GeV}, \beta=0.868950, \mu=8415.30 \,\mathrm{GeV}$. The mass of the heaviest scalar particle is 45.83 TeV. The divergences appear when an intermediate particle goes on shell in the $s$-channel. The grayed zone is not taken into account in the scans.}
\label{Fig: sqrtsscan}
\end{center}
\end{figure}

Fig.~\ref{Fig: sqrtsscan} shows the evolution of the largest eigenvalue of $\mathcal{T}^0$ at tree-level in function of $\sqrt{s}$. Due to the large number of scalars in our model, the amplitudes have numerous s-channel poles.
To avoid running into $s$, $t$ and $u$-channel singularities of the tree-level amplitudes, we follow the procedures outlined in Sec.~\ref{SubSec: poles}. 

The value of $\sqrt{s}$ that maximizes the $\mathcal{T}^0$ eigenvalues is in the result of a compromise between including more entries in the scattering matrix ---being above all the thresholds--- and having large $s$, $t$ and $u$-channel contributions. The resulting optimal value generally lies right behind the threshold corresponding to the heaviest scalar pair of the model
\begin{align}
    \sqrt{s_{\mathrm{threshold}}} = 2 m_{\mathrm{heaviest}}.
\end{align}
We found that computing unitarity constraints at
\begin{align}
    \sqrt{s_{\mathrm{max}}} = \sqrt{5} m_{\mathrm{heavy}}
\end{align}
constrains the parameter space as well as actually maximizing $\mathcal{T}^0_{ii}$ over $s$.
The optimal $\sqrt{s}$ is them of the same order as the heaviest scalar mass of the model. In Sec.~\ref{Sec: NMSSM}, we showed that, in the heavy limit, the NMSSM scalar Higgs masses are combinations of the scales $\mu$, $\sqrt{A_\lambda\mu}$ and $\sqrt{A_\kappa\mu}$. By constraining the ratios $A_\lambda^2/s$ and $A_\kappa^2/s$, unitarity bounds at $s = s_{\mathrm{max}}$ then effectively constrain the ratios $A_\lambda/\mu$ and $A_\kappa/\mu$.

Perturbativ unitarity constraints allow us to set upper bounds on the dimensionless couplings $\lambda$ and $\kappa$, but also on the ratios $A_\lambda/\mu$ and $A_\kappa/\mu$. Setting an upper bound on $\mu$ would then bound all the scales in the theory. Since $\mu$ appears only in the particle mass terms, an additional constraint other than unitarity is required to fully anchor the mass spectrum of the NMSSM Higgs sector. 
  
\section{Constraints on Thermal Dark Matter}
\label{Sec: Relic}
In order to set an upper bound on the mass scales $A_\lambda$, $A_\kappa$ and $\mu$, our perturbative unitarity constraints need to be supplemented by an additional requirement.  We require the NMSSM Higgs sector to have %
thermal Dark Matter candidate, which is in general a mixed Higgsino/singlino state. 
In the following, we require DM relic density to be smaller than or equal to the current measured value, equation~\leqn{eq:relic}.

It is possible that the observed dark matter in the universe is multi-component and is composed of, e.g., non-thermal axions.  Multiple contributions to the DM relic abundance means the NMSSM neutralinos must annihilate more efficiently so the total relic abundance matches equation~\leqn{eq:relic}.  This larger annihilation cross section means larger couplings, if all of the other parameters in the cross section are the same.  Thus, our perturbative unitarity bounds would be violated at lower scales than the ones presented in this paper.  See~\cite{Walker:2013hka} as an example of this for a non-supersymmetric Higgs portal.

\subsection{Unitarity and Relic Abundance}
In order to reach a low relic density, DM needs to efficiently annihilate to SM particles. The annihilation channels in the MSSM are shown in Fig.~\ref{Fig: MSSMann}. There involve only SM weak couplings. The higher the DM mass is, however, the more it needs to annihilate in order to have a low enough relic density. Since the couplings of the diagrams in Fig.~\ref{Fig: MSSMann} are fixed in the MSSM, there is an upper bound on how heavy the DM can be. This upper bound is around $1.1\,\mathrm{TeV}$ for MSSM Higgsinos \cite{Cirelli:2007xd}. 

In the NMSSM, however, additional annihilation channels, shown in Fig.~\ref{Fig: NMSSMann}, open for Higgsino/singlino DM. The couplings associated to these channels are proportional to $\lambda$ and $\kappa$. These couplings are not fixed and can therefore be large enough to allow multi-TeV DM particles to have the correct relic density. Since dimensionless couplings have to remain perturbative, a general model-independent upper bound on the DM mass can still be set by requiring all couplings to be less then $4\pi$. This bound, computed in \cite{Griest:1989wd}, is of about $120\,\mathrm{TeV}$. In our study, however, the unitarity criteria detailed in Sec.~\ref{Sec: Unitarity} set much tighter bounds on $\lambda$ and $\kappa$. Using these bounds would then allow to significantly improve the bound on the DM mass set in \cite{Griest:1989wd}.

As shown in Sec.~\ref{Sec: NMSSM}, the DM mass in the heavy limit is $\mathcal{O}(\mu)$. Setting a bound on the DM mass using relic density would then amount to set an upper bound on $\mu$. Using the unitarity constraints from Sec.~\ref{Sec: Unitarity} would then allow to fully anchor the mass spectrum in our model.
\begin{figure}[!h]
    \centering
    \begin{fmffile}{MSSMW}
        \begin{fmfgraph*}(150,130)
            \fmfset{arrow_len}{2.5mm}
            \fmfset{arrow_ang}{25}
            \fmfset{wiggly_len}{5mm}
            \fmfset{wiggly_slope}{70}
            \fmfleft{i1,i2}
            \fmfright{o1,o2}
            \fmf{fermion, label=$\tilde{\chi}$,label.side=left}{i1,v1}
            \fmf{photon,tension=0.1}{i1,v1}
            \fmf{photon,tension=0.1}{i2,v2}
            \fmf{fermion, label=$\tilde{\bar{\chi}}$,label.side=right}{i2,v2}
            \fmf{fermion, label=$\tilde{\chi}^\pm$,tension=0.5}{v2,v1}
            \fmf{photon, label=$W^\pm$,label.side=left}{v1,o1}
            \fmf{photon, label=$W^\mp$}{v2,o2}
        \end{fmfgraph*}
    \end{fmffile}\quad\quad
    \begin{fmffile}{MSSMZ}
        \begin{fmfgraph*}(150,130)
            \fmfset{arrow_len}{2.5mm}
            \fmfset{arrow_ang}{25}
            \fmfset{wiggly_len}{5mm}
            \fmfset{wiggly_slope}{70}
            \fmfleft{i1,i2}
            \fmfright{o1,o2}
            \fmf{fermion, label=$\tilde{\chi}$,label.side=left}{i1,v1}
            \fmf{photon,tension=0.1}{i1,v1}
            \fmf{photon,tension=0.1}{i2,v2}
            \fmf{fermion, label=$\tilde{\bar{\chi}}$,label.side=right}{i2,v2}
            \fmf{fermion, label=$\tilde{\chi}_i^0$,tension=0.5}{v2,v1}
            \fmf{photon, label=$Z$,label.side=left}{v1,o1}
            \fmf{photon, label=$Z$}{v2,o2}
        \end{fmfgraph*}
    \end{fmffile}
    \caption{\label{Fig: MSSMann} Dark Matter annihilation diagrams to gauge bosons through a chargino/neutralino. The corresponding amplitudes are proportional to the SM weak couplings.}
\end{figure}
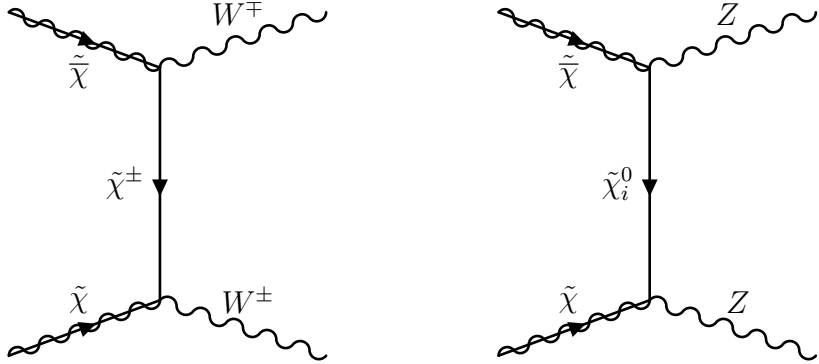
\begin{figure}[!h]
    \centering
    \begin{fmffile}{tpartner2}
        \begin{fmfgraph*}(200,120)
            \fmfset{arrow_len}{2.5mm}
            \fmfset{arrow_ang}{25}
            \fmfset{wiggly_len}{5mm}
            \fmfset{wiggly_slope}{70}
            \fmfleft{i1,i2}
            \fmfright{o1,o2}
            \fmf{fermion, label=$\tilde{\chi}$,label.side=left}{i1,v1}
            \fmf{photon,tension=0.1}{i1,v1}
            \fmf{photon,tension=0.1}{i2,v1}
            \fmf{fermion, label=$\tilde{\bar{\chi}}$,label.side=right}{i2,v1}
            \fmf{dashes, label=$h_{123},,a_{12}$,tension=0.5}{v1,v2}
            \fmf{fermion, label=SM,label.side=left}{v2,o1}
            \fmf{fermion, label=SM}{v2,o2}
        \end{fmfgraph*}
    \end{fmffile}
    \caption{\label{Fig: NMSSMann} s-channel Dark Matter annihilation diagrams through scalar and pseudoscalar Higgs bosons in the NMSSM. The corresponding amplitudes are proportional to a combination of the $\lambda$ and $\kappa$ quartic couplings.}
\end{figure}
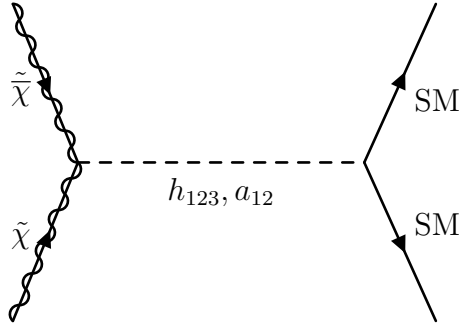

\subsection{Possible loopholes}
Combining unitarity and relic density constraints allows to set upper bounds on the $A_{\lambda\kappa}$ and $\mu$ scales for most of the parameter configurations. In some regions of the parameter space, however, the DM annihilation rate is strongly enhanced even when the $\lambda$ and $\kappa$ couplings are small. In these regions, the bounds set by unitarity and relic density will be considerably loosened. These regions are, however, narrow and well-defined parts of the parameter space, resulting from a significant fine-tuning of the NMSSM parameters.
\paragraph{$s$-channel resonances}
If the DM mass is half the mass of one of the neutral Higgses, the annihilation cross section will receive large contributions from $s$-channel resonant diagrams. To quantify how close one is to the s-channel resonant regions, one can use the fine-tuning parameter $R_i$, defined as
\begin{align}
    R &= \mathrm{max}_i \frac{\left|2m_{\mathrm{DM}} - m_{H_i}\right|}{m_{H_i}}.
    \label{Eq: Rparam}
\end{align}
$R$ is positive and goes to zero in the Higgs funnel regions. Our bounds will not apply in parameter regions with small value of $R$.

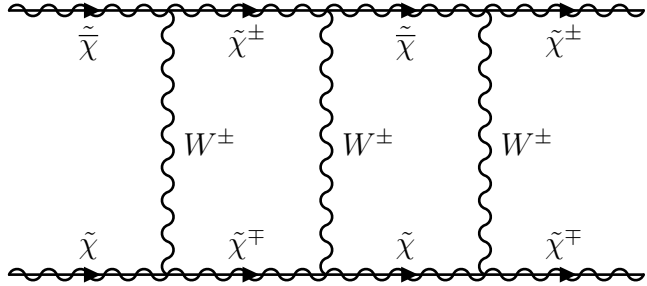
\begin{figure}[!h]
    \centering
    \begin{fmffile}{sommerfeld}
        \begin{fmfgraph*}(300,100)
            \fmfset{arrow_len}{2.5mm}
            \fmfset{arrow_ang}{25}
            \fmfset{wiggly_len}{5mm}
            \fmfset{wiggly_slope}{70}
            \fmfleft{i1,i2}
            \fmfright{o1,o2}
            \fmf{fermion, label=$\tilde{\chi}$,label.side=left}{i1,v1}
            \fmf{photon,tension=0.1}{i1,v1}
            \fmf{photon,tension=0.1}{i2,v2}
            \fmf{fermion, label=$\tilde{\bar{\chi}}$,label.side=right}{i2,v2}
            \fmf{photon,label=$W^\pm$,tension=0}{v1,v2}
            \fmf{fermion, label=$\tilde{\chi}^\mp$,label.side=left}{v1,v3}
            \fmf{fermion, label=$\tilde{\chi}^\pm$}{v2,v4}
            \fmf{photon,tension=0.1}{v1,v3}
            \fmf{photon,tension=0.1}{v2,v4}
            \fmf{photon,label=$W^\pm$,tension=0}{v3,v4}
            \fmf{fermion, label=$\tilde{\chi}$,label.side=left}{v3,v5}
            \fmf{fermion, label=$\tilde{\bar\chi}$}{v4,v6}
            \fmf{photon,tension=0.1}{v3,v5}
            \fmf{photon,tension=0.1}{v4,v6}
            \fmf{photon,label=$W^\pm$,tension=0}{v5,v6}
            \fmf{fermion, label=$\tilde{\chi}^\mp$,label.side=left}{v5,o1}
            \fmf{fermion, label=$\tilde{\chi}^\pm$}{v6,o2}
            \fmf{photon,tension=0.1}{v5,o1}
            \fmf{photon,tension=0.1}{v6,o2}
        \end{fmfgraph*}
    \end{fmffile}
    \caption{\label{Fig: sommerfeld} Example of a ladder diagram contributing to the Sommerfeld enhancement for Higgsino DM annihilation.}
\end{figure}

\paragraph{$t$-channel resonances}
In both the NMSSM and the MSSM, DM can annihilate in the $t$-channel to two $W$ bosons with an intermediate chargino. For
\begin{align}
    m_{\mathrm{DM}} - m_{\mathrm{Chargino}} \sim m_{W^\pm}
\end{align}
resonant annihilation occurs. In our model, the neutralino mass is either smaller than or equal to the chargino mass and so such $t$-channel resonant annihilations never take place.
\paragraph{Sommerfeld enhancement}
In certain cases, the annihilating DM particles form bound states before annihilating. Such bound states are formed through ladder diagrams like the one shown in Fig.~\ref{Fig: sommerfeld}. This non-perturbative process can significantly enhance the DM annihilation cross section and relax the bounds set by relic density. In the MSSM for example, it increases the upper bound on the wino mass by a factor of two \cite{Cirelli:2007xd}. The magnitude of the Sommerfeld enhancement factor depends on the sizes of the following parameters
        \begin{align}
		\epsilon_v &= {1\over\alpha}\left({v\over c}\right),\quad
                \epsilon_{\delta} = {1\over\alpha} \sqrt{2\delta\over m_{\mathrm{DM}}},\quad
		\epsilon_{\phi} = {1\over\alpha} \left ({m_{\phi}\over m_{\chi}}\right)
        \end{align}
        where $v$ is the DM velocity, $\alpha$ is the fine-structure constant of the interaction at play in the ladder diagram and $m_\phi$ is the mass of the mediating particles ---the $W$ bosons in Fig.~\ref{Fig: sommerfeld}. $\delta$ is the mass splitting between DM and the intermediate particle in the ladder diagram ---so the chargino in Fig.~\ref{Fig: sommerfeld}. The Sommerfeld enhancement is significant only if $\epsilon_v,\epsilon_\delta, \epsilon_\phi \lesssim 1$. For multi-TeV DM, $\epsilon_v, \epsilon_\phi \sim 0$. The size of the Sommerfeld enhancement will then depend on the Higgsino-Singlino or the Higgsino-Chargino mass splittings. The former is larger than a few GeV at tree-level in most of the parameter space. The latter usually receives large one-loop contributions from stops and sbottoms in the full NMSSM. Therefore, in most of the parameter space, $\epsilon_\delta\gsim 1$ and Sommerfeld enhancement can be neglected. 

\subsection{Direct Detection}
\begin{figure}[!h]
    \centering
    \begin{fmffile}{directdetection}
        \begin{fmfgraph*}(200,120)
            \fmfset{arrow_len}{2mm}
            \fmfset{arrow_ang}{20}
            \fmfset{wiggly_len}{6mm}
            \fmfset{wiggly_slope}{80}
            \fmfleft{i1,i2} 
            \fmfright{o1,o2} 
            \fmf{fermion,label=$\tilde{\bar \chi}$,label.side=left}{i1,v1}
            \fmf{fermion,label=$\tilde{\chi}$}{v1,i2}
            \fmf{dashes}{v1,v2}
            \fmf{fermion}{o1,v2,o2}
            \fmffreeze
            \fmfblob{.15w}{v2}
            \fmfi{plain}{vpath (__o1,__v2) shifted (thick*(2,1))} 
            \fmfi{plain}{vpath (__o1,__v2) shifted (thick*(-2.5,0))}
            \fmfi{plain}{vpath (__v2,__o2) shifted (thick*(2,-1))} 
            \fmfi{plain}{vpath (__v2,__o2) shifted (thick*(-2.5,0))}
        \end{fmfgraph*}
    \end{fmffile}
    \caption{\label{Fig: SIdirect} Tree-level diagrams contributing to the spin-independent direct detection cross-section for Higgsino/Singlino DM. The associated cross sections are mixing angle suppressed.}
\end{figure}
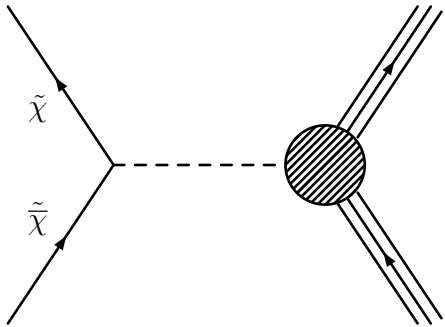

In the MSSM, there is no tree-level contribution to the spin-independent direct detection (DD) cross section for Higgsino Dark Matter. In the NMSSM, however, Higgsino/Singlino states can scatter against up and down quarks through the diagrams shown in Fig.~\ref{Fig: SIdirect}, which give a non-zero spin-independent DD cross section. For heavy DM, these diagrams are strongly suppressed by the sine of the mixing angle between the doublet and singlet scalar states. Future direct detection experiments like XENON1T, though, would still be able to reach part of the light DM regions. 

\section{Results}
\label{Sec: Results}
\begin{figure}
\begin{center}
\vskip -0.3in
\includegraphics[height=3in]{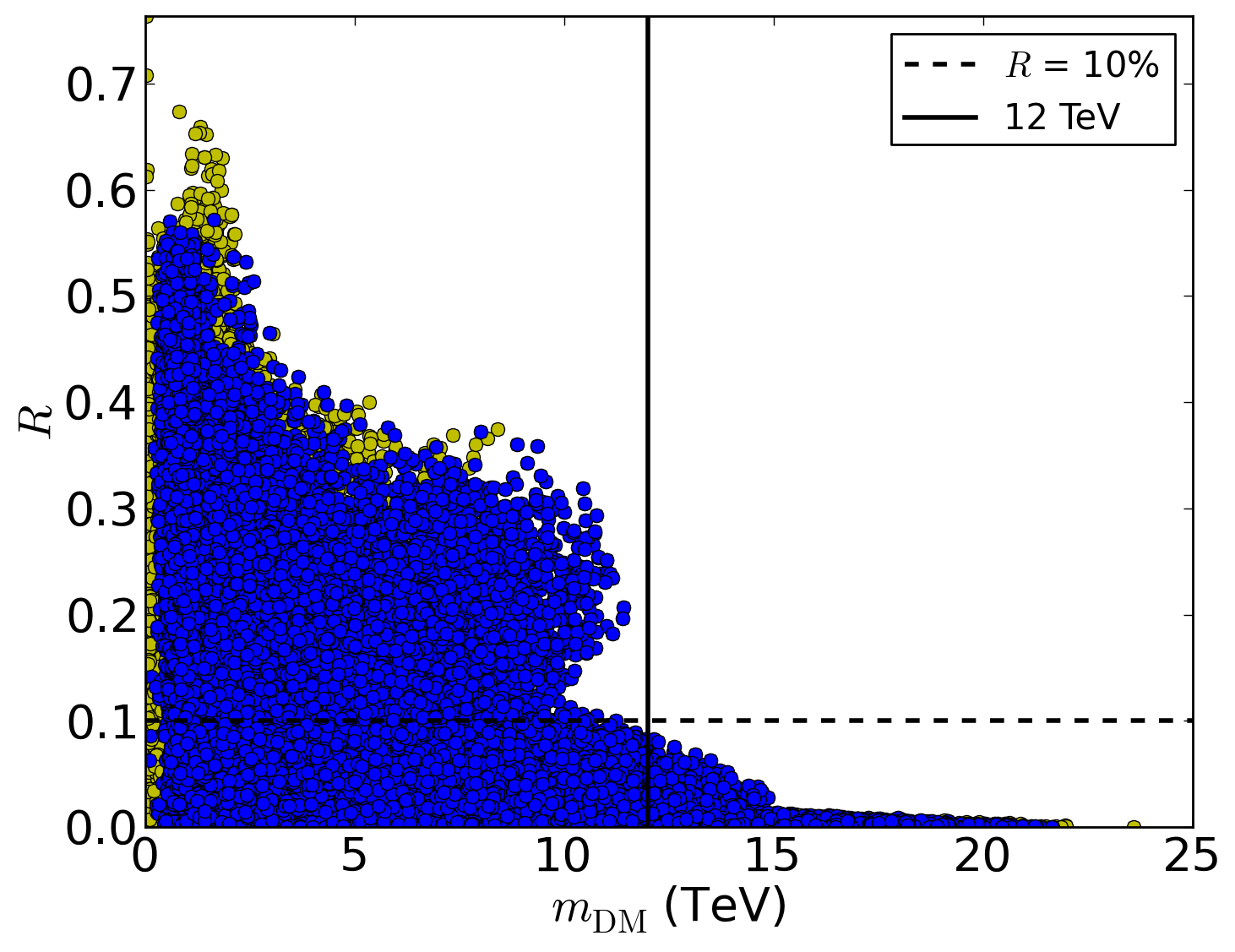}
\caption{$R$ fine-tuning factor versus the DM mass for the points passing the unitarity and relic density constraints for $a_{\mathrm{max}} = 41\%$. The yellow points represents points that are within the reach of XENON1T while the blue points are expected not to be seen by the next generation of DM direct detection experiments.}
\label{Fig: mdm}
\end{center}
\end{figure}

\begin{figure}
\begin{center}
\vskip -0.3in
\includegraphics[height=3in]{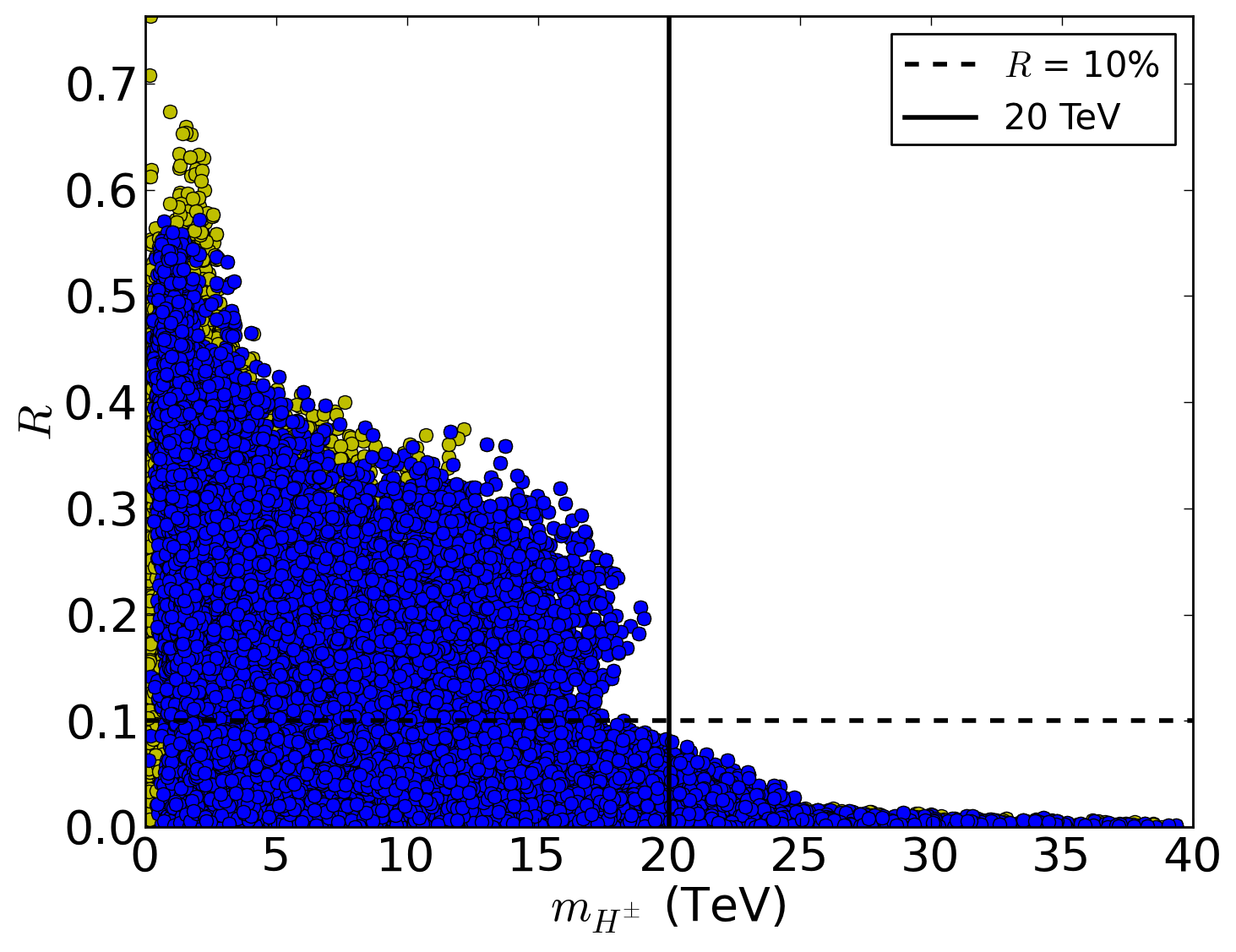}
\caption{$R$ fine-tuning factor versus the mass of the charged Higgs for the points passing the unitarity and relic density constraints for $a_{\mathrm{max}} = 41\%$. The yellow points represents points that are within the reach of XENON1T while the blue points are expected not to be seen by the next generation of DM direct detection experiments.}
\label{Fig: mch}
\end{center}
\end{figure}

\begin{figure}
\begin{center}
\vskip -0.3in
\includegraphics[height=3in]{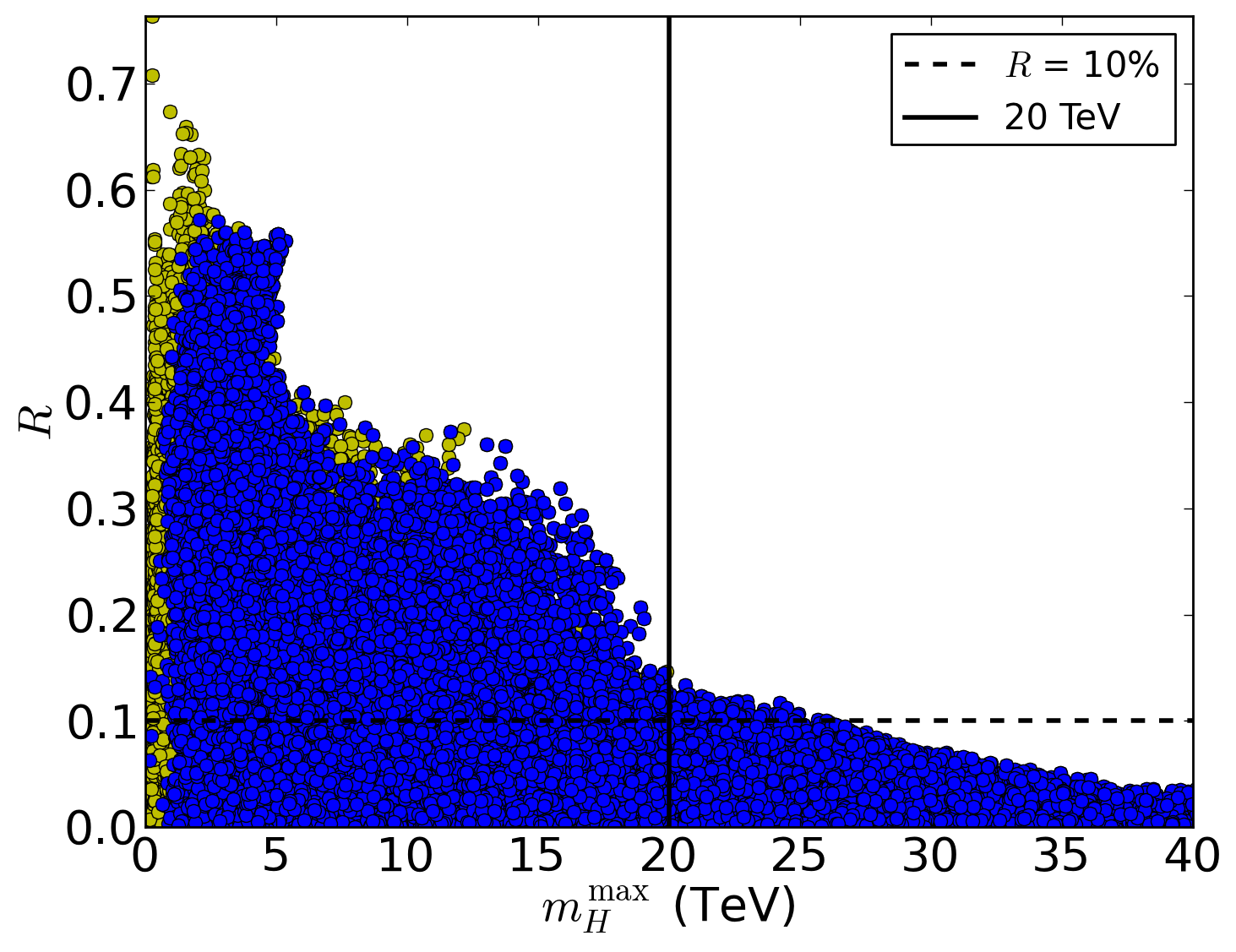}
\caption{$R$ fine-tuning factor versus the heaviest CP-even Higgs mass for the points passing the unitarity and relic density constraints for $a_{\mathrm{max}} = 41\%$. The yellow points represents points that are within the reach of XENON1T while the blue points are expected not to be seen by the next generation of DM direct detection experiments.}
\label{Fig: mhe}
\end{center}
\end{figure}

\begin{figure}
\begin{center}
\vskip -0.3in
\includegraphics[height=3in]{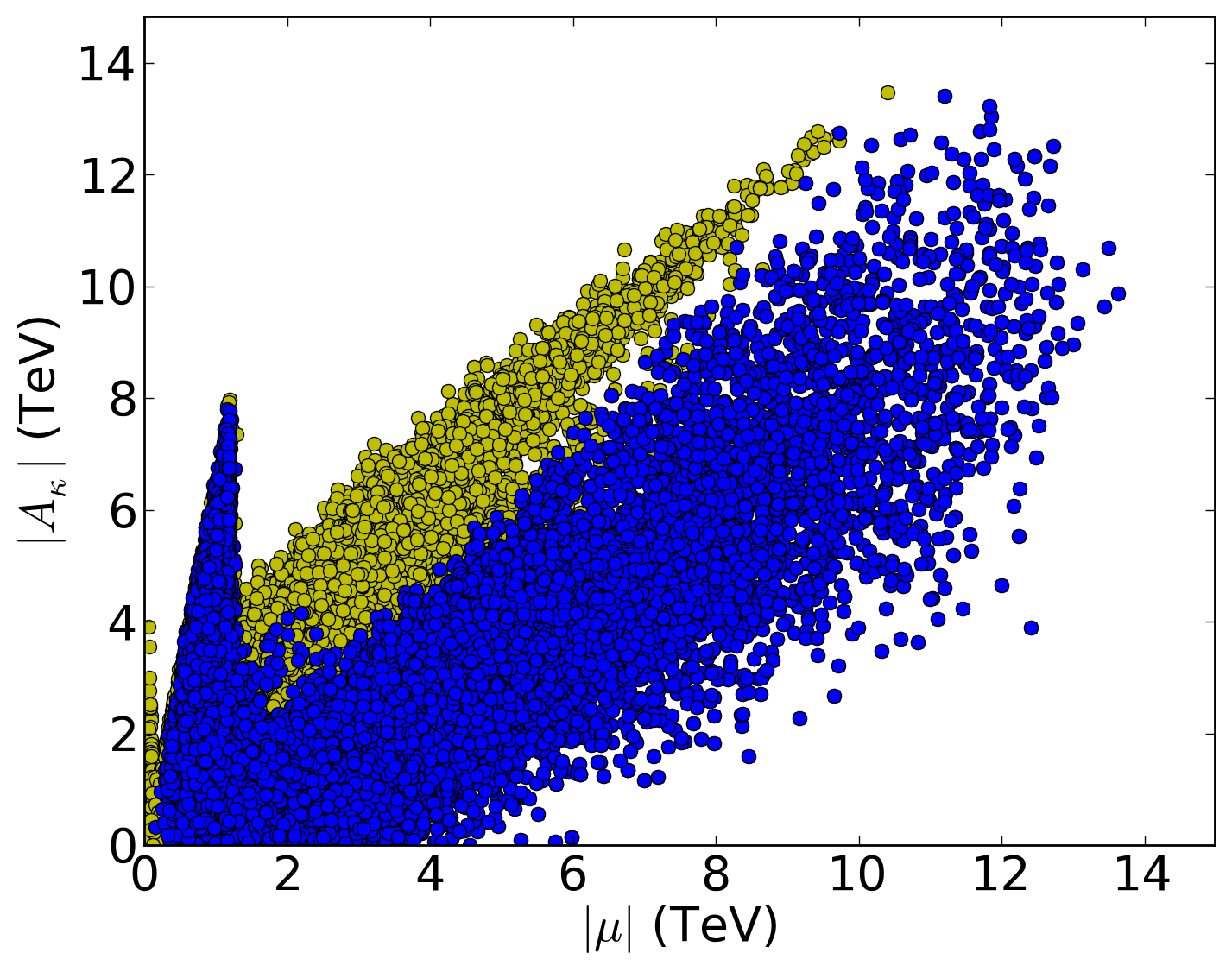}
\caption{$|A_\kappa|$ versus $|\mu|$ for the non-resonant points passing the unitarity and relic density constraints for $a_{\mathrm{max}} = 41\%$. The yellow points represents points that are within the reach of XENON1T while the blue points are expected not to be seen by the next generation of DM direct detection experiments.}
\label{Fig: akmu}
\end{center}
\end{figure}

\begin{figure}
\begin{center}
\vskip -0.3in
\includegraphics[height=3in]{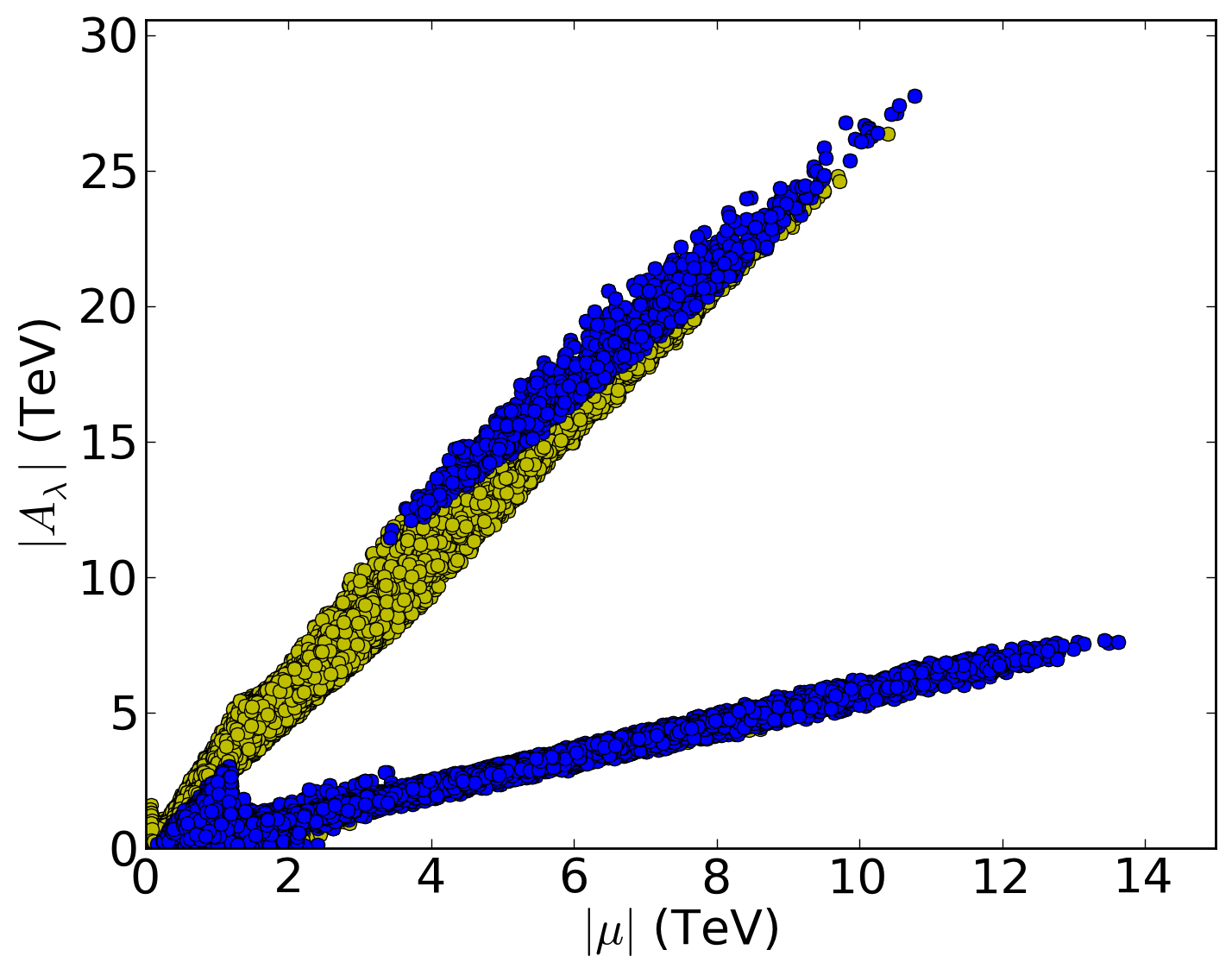}
\caption{$|A_\lambda|$ versus $|\mu|$ for the non-resonant points passing the unitarity and relic density constraints for $a_{\mathrm{max}} = 41\%$. The yellow points represents points that are within the reach of XENON1T while the blue points are expected not to be seen by the next generation of DM direct detection experiments.}
\label{Fig: almu}
\end{center}
\end{figure}

We scan uniformly over the five-dimensional NMSSM parameter space using the following scan bounds
\begin{align}
    |\lambda|, |\kappa| \leq 4, \quad\text{and}\quad |\mu|, |A_\lambda| < 40\,\mathrm{TeV}.
\end{align}
$A_\kappa$ is fixed by requiring one of the Higgs mass eigenstates to be at $125\,\mathrm{GeV}$. We then select points of the parameter space using the following requirements
\begin{itemize}
    \item No tachyonic masses
    \item Stable EWSB vacuum
    \item Unitary scattering amplitudes
    \item Relic density lower than the current measured value
    \item Direct detection cross section lower than the current LUX bounds \cite{Akerib:2013tjd}.
\end{itemize}
The unitarity bounds are computed for both $a_{\max} = 41\%$ and $a_{\max} = 20\%$. The tree-level scattering amplitudes are evaluated at $\sqrt{s} = \sqrt{5} \,m_{\mathrm{Heaviest}}$. Relic densities and spin-independent direct detection cross sections are computed using MicrOmegas \cite{Belanger:2013oya}. The maximal value of the relic density is taken to be the value measured by Planck \cite{Ade:2013zuv} plus three sigma
\begin{align}
    \Omega_{\mathrm{max}} \le 0.1199 \pm 0.0027
\end{align}
Figs \ref{Fig: mdm}, \ref{Fig: mch}, \ref{Fig: mhe}, \ref{Fig: akmu} and \ref{Fig: almu} show the points of the parameter space that survive all the cuts for $a_{\mathrm{max}} = 41\%$. These figures also show the projected reach of the XENON1T experiment. The yellow points are points that would be within the reach of XENON1T while the blue points will be outside the reach of the experiment. In order to separate between the bulk of the parameter space and the s-channel resonant regions, Figs \ref{Fig: mdm} to \ref{Fig: mhe} show the fine-tuning parameter $R$, defined in ~\leqn{Eq: Rparam} versus the masses of the DM, the charged Higgs and the heaviest CP-even Higgs respectively. Outside the resonant region, the bound on the DM mass is of about $12\,\mathrm{TeV}$, so one order of magnitude larger than the bound shown in \cite{Griest:1989wd}. The upper bounds on the charged Higgs mass and the heaviest CP-even Higgs mass are about a factor of two larger, around $20$ and $25\,\mathrm{TeV}$ respectively.

 Figs \ref{Fig: akmu} and \ref{Fig: almu} respectively show $|A_\lambda|$ versus $|\mu|$ and $|A_\kappa|$ versus $|\mu|$ for points outside the resonant regions. Points are considered outside the resonant regions if
 \begin{align}
     R > 10\%.
 \end{align}
 Fig.~\ref{Fig: almu}, that shows $|A_\lambda|$ versus $|\mu|$ is particularly striking. Here, vacuum constraints favors regions of the parameter space where the ratio $A_\lambda/\mu$ is well-defined. With the $a_{\mathrm{max}} = 41\%$ unitarity criterium, typical $A_\lambda/\mu$ ratios are no larger than about $2$.

\begin{figure}
\begin{center}
\vskip -0.3in
\includegraphics[height=3in]{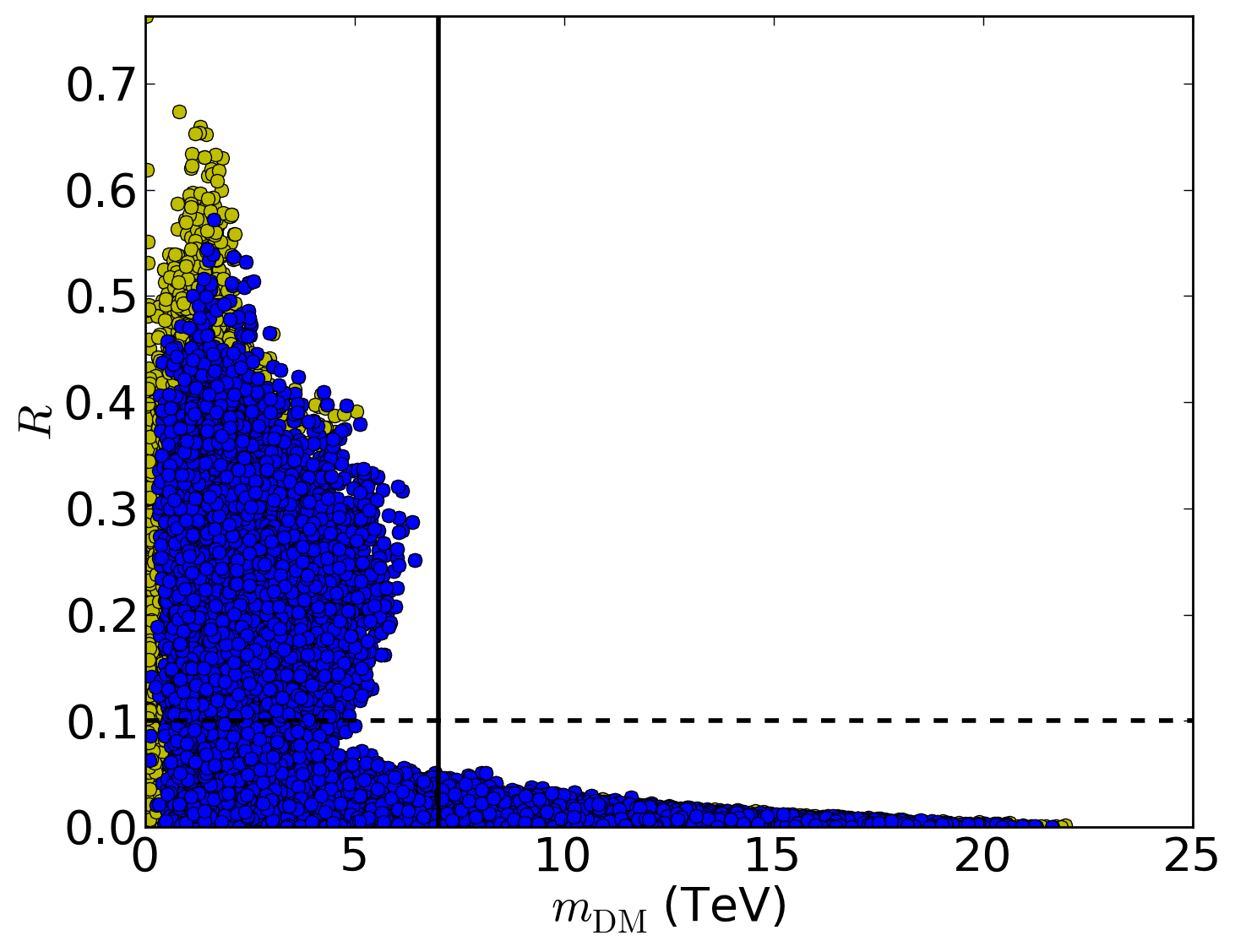}
\caption{$R$ fine-tuning factor versus the DM mass for the points passing the unitarity and relic density constraints for $a_{\mathrm{max}} = 20\%$. The yellow points represents points that are within the reach of XENON1T while the blue points are expected not to be seen by the next generation of DM direct detection experiments.}
\label{fig:mdm}
\end{center}
\end{figure}

\begin{figure}
\begin{center}
\vskip -0.5in
\includegraphics[height=3in]{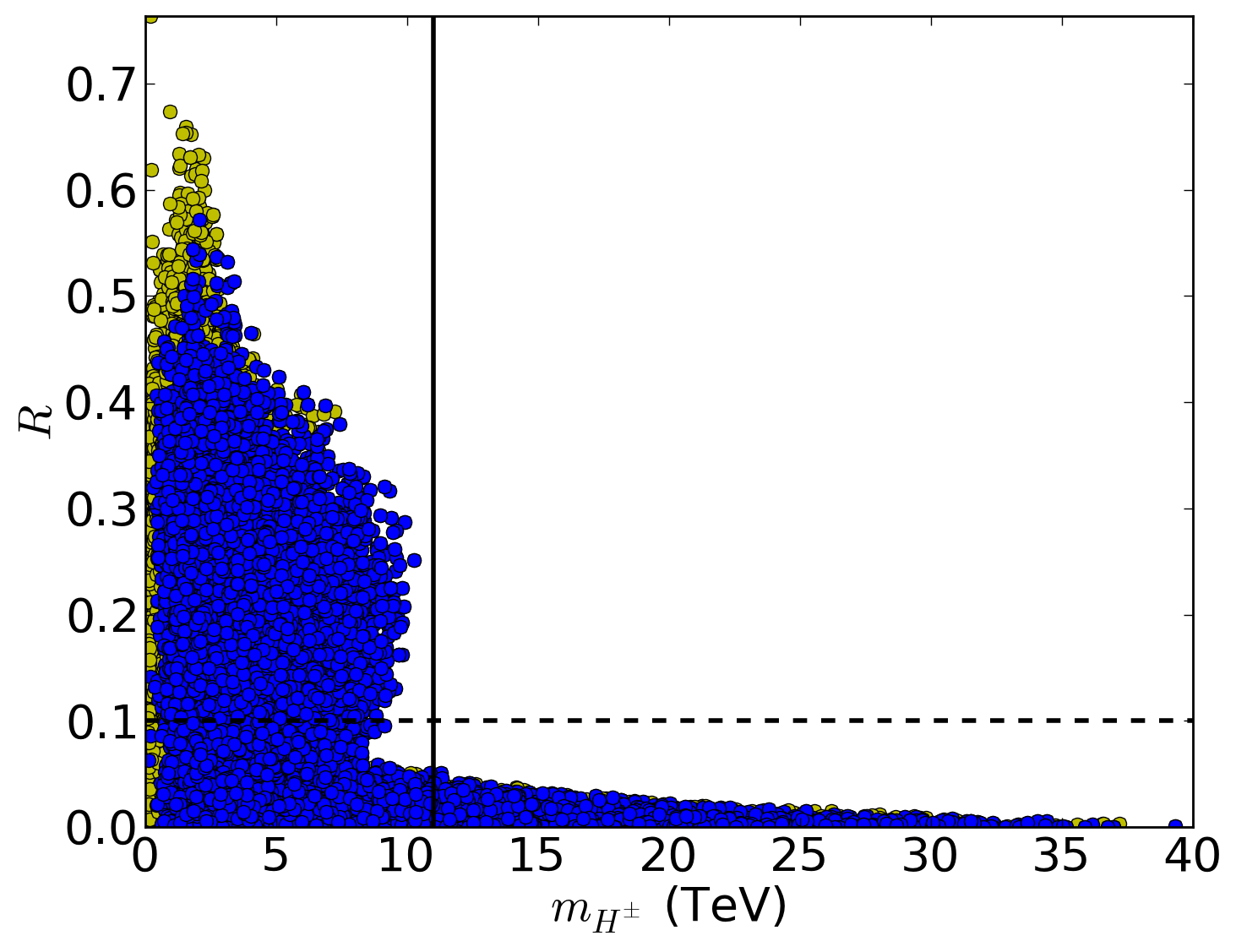}
\caption{$R$ fine-tuning factor versus the charged Higgs mass for the points passing the unitarity and relic density constraints for $a_{\mathrm{max}} = 20\%$. The yellow points represents points that are within the reach of XENON1T while the blue points are expected not to be seen by the next generation of DM direct detection experiments.}
\label{fig:mch}
\end{center}
\end{figure}

\begin{figure}
\begin{center}
\vskip -0.5in
\includegraphics[height=3in]{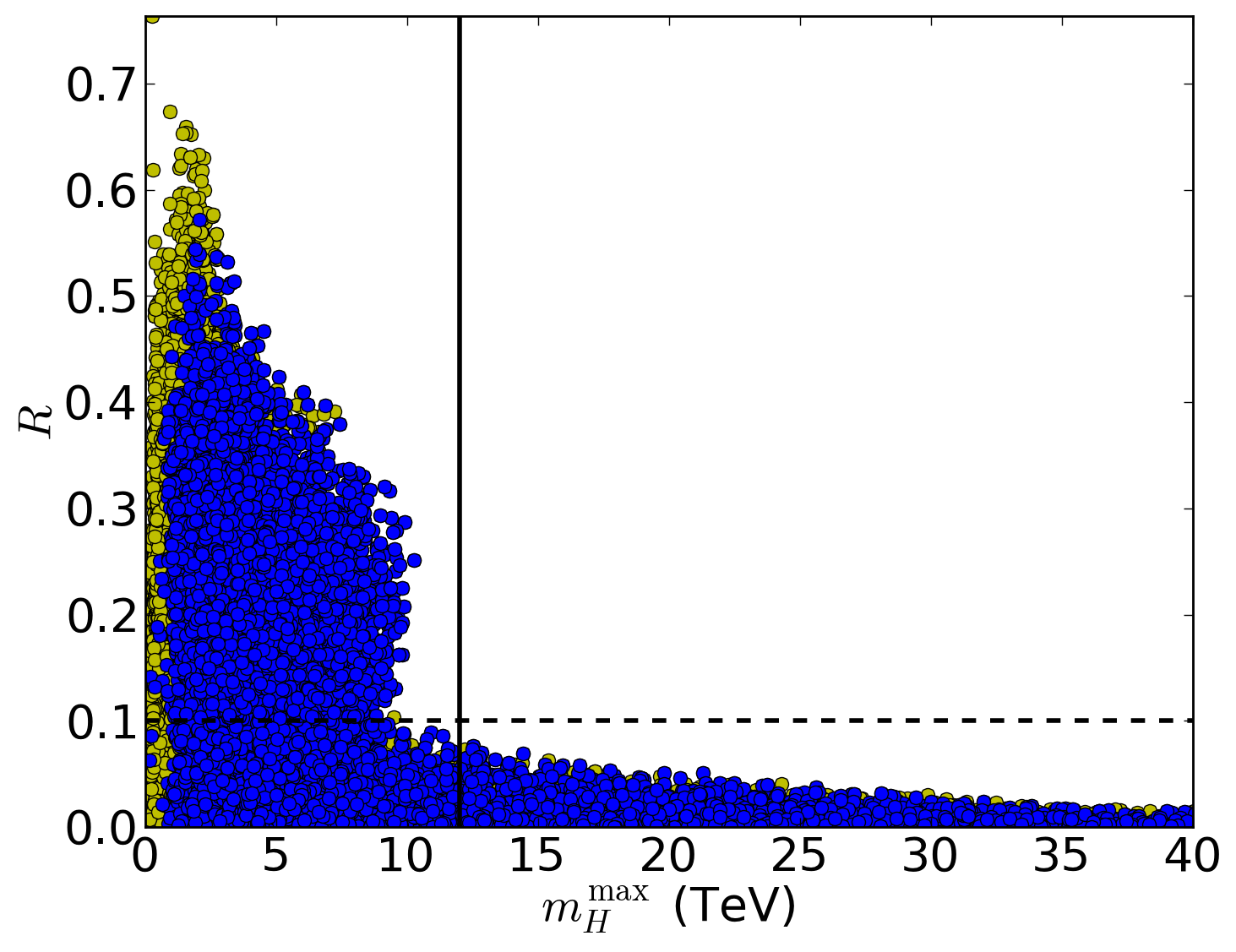}
\caption{$R$ fine-tuning factor versus the heaviest CP-even Higgs mass for the points passing the unitarity and relic density constraints for $a_{\mathrm{max}} = 20\%$. The yellow points represents points that are within the reach of XENON1T while the blue points are expected not to be seen by the next generation of DM direct detection experiments.}
\label{fig:mhe}
\end{center}
\end{figure}

\begin{figure}
\begin{center}
\vskip -0.3in
\includegraphics[height=3in]{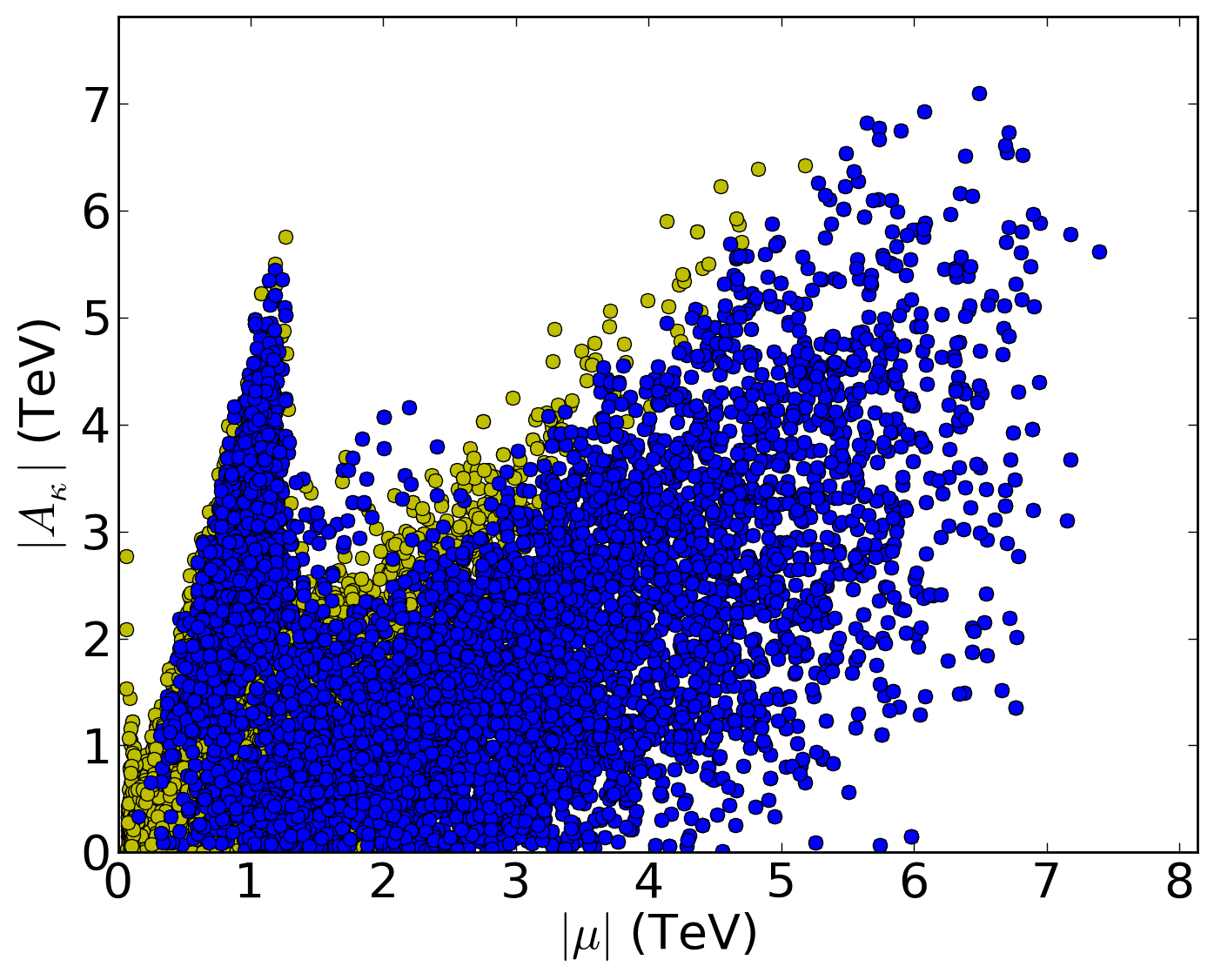}
\caption{$|A_\kappa|$ versus $|\mu|$ for the non-resonant points passing the unitarity and relic density constraints for $a_{\mathrm{max}} = 41\%$. The yellow points represents points that are within the reach of XENON1T while the blue points are expected not to be seen by the next generation of DM direct detection experiments.}
\label{fig:akmu}
\end{center}
\end{figure}

\begin{figure}
\begin{center}
\vskip -0.3in
\includegraphics[height=3in]{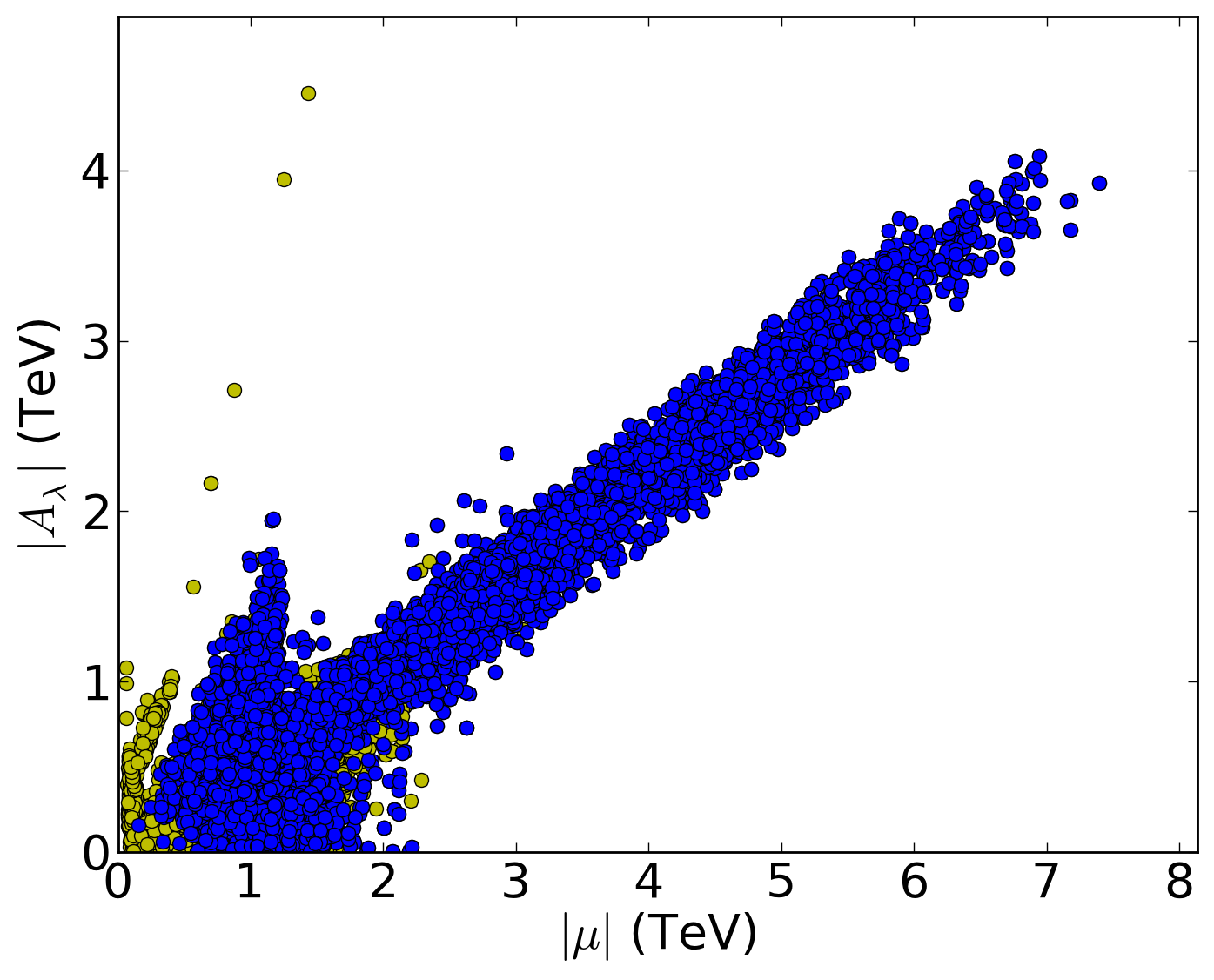}
\caption{$|A_\lambda|$ versus $|\mu|$ for the non-resonant points passing the unitarity and relic density constraints for $a_{\mathrm{max}} = 41\%$. The yellow points represents points that are within the reach of XENON1T while the blue points are expected not to be seen by the next generation of DM direct detection experiments.}
\label{fig:almu}
\end{center}
\end{figure}

Figs \ref{fig:mdm}, \ref{fig:mch}, \ref{fig:mhe}, \ref{fig:akmu} and \ref{fig:almu} show the same plots as Figs \ref{Fig: mdm} to \ref{Fig: almu} but for $a_{\mathrm{max}} = 20\%$. Figs \ref{fig:mdm} to \ref{fig:mhe} show $R$ versus the DM, charged Higgs and heaviest CP-even Higgs masses respectively. Figs \ref{fig:akmu} and \ref{fig:almu} respectively show $|A_\lambda|$ versus $|\mu|$ and $|A_\kappa|$ versus $|\mu|$ for non-resonant points. The upper bound on the DM mass outside the resonant regions is now of about $7\,\mathrm{TeV}$, so tighter than with the $a_{\mathrm{max}} = 41\%$ unitarity criterium. With the new, tighter, unitarity criterium, the upper bound on the heaviest Higgs mass gets much stronger and is now of about $10\,\mathrm{TeV}$. 
 
\begin{figure}
\begin{center}
\vskip -0.3in
\includegraphics[height=3in]{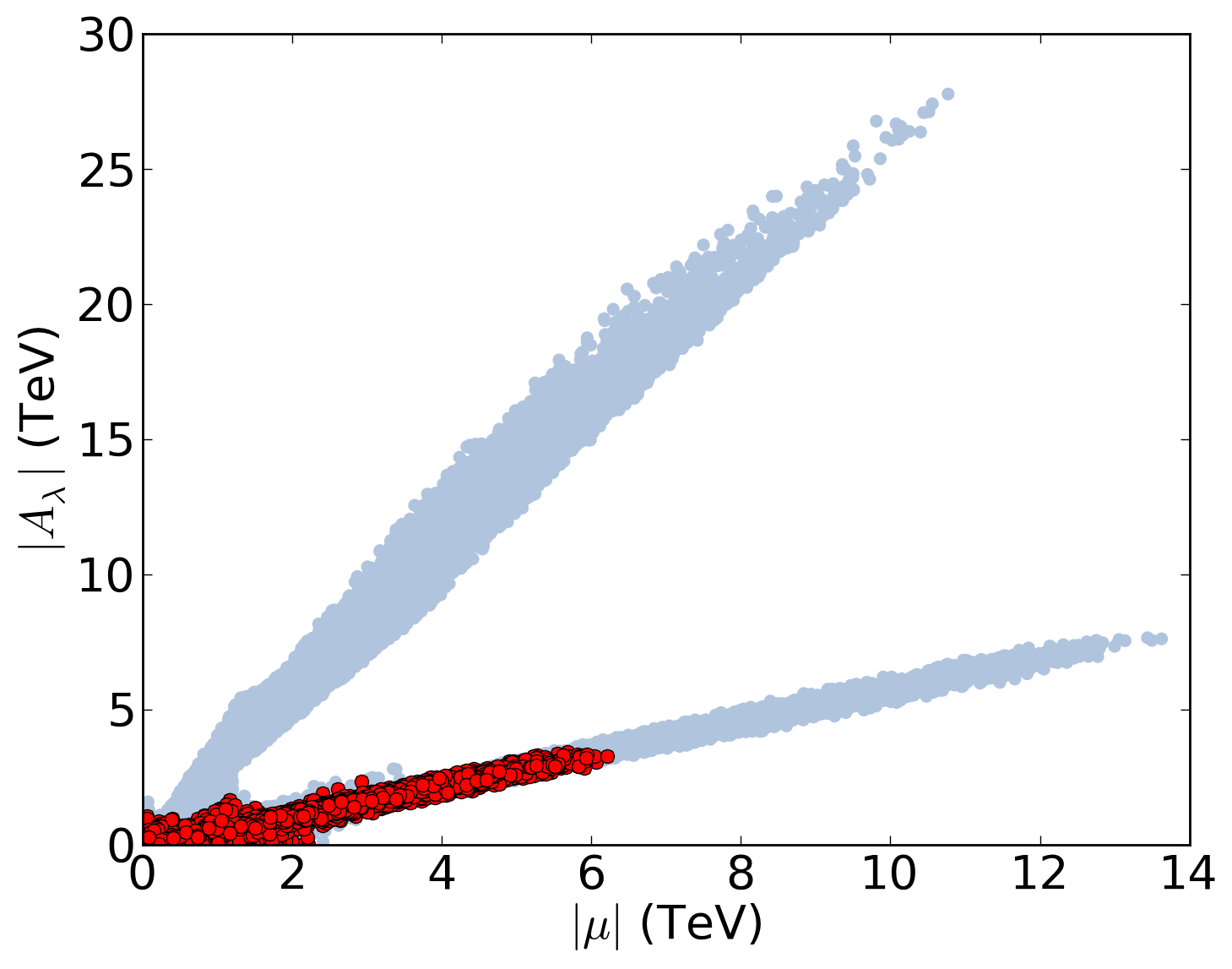}
\caption{$|A_\lambda|$ versus $|\mu|$ for the non-resonant points passing the unitarity and relic density constraints for $a_{\mathrm{max}} = 41\%$ (light blue) and $a_{\mathrm{max}} = 20\%$ (red). The large $\left|A_\lambda/\mu\right|$ region is ruled out when the unitarity constraint becomes tighter.}
\label{fig:compUnit}
\end{center}
\end{figure}

 This significant drop in the Higgs mass bound can be understood by noticing that the typical $A_\lambda/\mu$ ratios shown in Fig.~\ref{fig:almu} are much smaller than with $a_{\mathrm{max}} = 41\%$. Fig.~\ref{fig:compUnit} shows $|A_\lambda|$ versus $|\mu|$ for both $a_{\mathrm{max}} = 41\%$ and $a_{\mathrm{max}} = 20\%$. Tightening the unitarity criterium has caused the large $A_\lambda/\mu$ branch to disappear, leaving only points with $A_\lambda/\mu \sim 1$. Fig.~\ref{fig:compUnit} illustrates how unitarity criteria can be used to constrain ratios of energy scales.



\pagebreak

\section{Conclusion}
\label{Sec: Conclusion}
We have seen that vacuum stability together with perturbative unitarity and relic abundance considerations put powerful constraints on the scale of new physics in the NMSSM for the majority of the parameter space. The points that are unconstrained by these considerations are finely tuned in the sense that there is a resonance decay mechanism or large Sommerfeld enhancement facilitated by small mass splitting between the LSP and the charginos. 

Outside of the finely tuned regions, next generation collider, direct and indirect detection experiments will be able to eliminate most of the parameter space leading us to very special finely tuned parameter points within the NMSSM.

\Acknowledgements
We thank M.~Cahill-Rowley, J.~Donoghue, A.~Nelson, M.~Peskin, T.~Rizzo and D.~Zeppenfeld for useful discussions.  We thank C.~Berger for accurate translations of~\cite{Schuessler:thesis} and M.~Peskin and T.~Rizzo reading drafts of this letter. This work is supported by the US Department of Energy, contract DE-ACO2-76SF00515. DW is also supported in part by a grant from the Ford Foundation via the National Academies of the Sciences as well as the National Science Foundation under Grants No. NSF PHY11-25915 and NSF-PHY-0705682.  SEH is supported by a Stanford Graduate Fellowship.

\pagebreak
\appendix

\section{Handling of Tree-level poles}
\subsection{s-channel poles}
s-channel propagators are of the following form
\begin{align}
    \mathcal{D} = \frac{1}{s - m^2 + i\Gamma}
\end{align}
where $\Gamma$ is the decay width of the propagating particle. In order for the tree-level approximation
\begin{align}
    \mathcal{D}^{\mathrm{tree}} = \frac{1}{s - m^2}
\end{align}
to be valid, the width $\Gamma$ needs to be small with respect to $s - m^2$. In this study, we follow the approach outlined in \cite{Schuessler:2007av} and require
\begin{align}
    \left|\sqrt{s} - m\right| > b \Gamma
\end{align}
where $b \gsim 1$ is a regulator. $b$ should be chosen such that the regulator dependence of the final result is minimal. In our study, we set $b$ to the value used in \cite{Schuessler:2007av} for the MSSM:
\begin{align}
    b = 3.3.
\end{align}
However, we found that the optimal value of $s$ that gives the best constraint while avoiding any $s-$channel poles lies close to the value $s=5m^2_{H_{\text{heaviest}}}$, where $m_{H_{\text{heaviest}}}$ is the mass of the heaviest scalar Higgs. We will discuss this further in Section~\ref{subsec:sFinite}.

\subsection{t and u-channel poles}
The partial wave components of the scattering matrix are obtained by integrating over the scattering angle $\theta$. Regions where a $t$ or $u$-channel pole is reached for at least one $\theta$ should then be treated with care. In particular, for a two-to-two scattering process of the form $1+2 \rightarrow 3+4$, $t$ and $u$-channel poles are encountered in the following configurations
	\ben
		\item $m_1>m_m+m_3$ and $m_2+m_m<m_4$ in $t$-channel scattering, 
		\item $m_1>m_m+m_4$ and $m_2+m_m<m_3$ in $u$-channel scattering.
                    \label{Eq: crit}
	\een
        Initial and final states that verify either of these conditions are called critical states. Critical states are states that are involved in at least one scattering process with a $t$ or $u$-channel pole. Scattering processes involving a critical state should then not be taken into account when diagonalizing the scattering matrix. In this study, we follow the partial diagonalization procedure followed in \cite{Schuessler:2007av} that is outlined in more details below\footnote{The rest of the section is taken directly from \cite{Schuessler:2007av}. Since the thesis outlining this procedure in detail is in German, we considered it would be useful to transcribe this part here.}.
        
        For two-to-two scattering processes, we divide the set of all possible 2-particle states into critical states $K$ and not-critical states $NK$ using the conditions shown in ~\leqn{Eq: crit}. For an given initial state $i$ and a final state $j$, both non-critical, ~\leqn{Eq: optic} becomes
        \beq
        \Im\,\mathcal{T}^J_{fi}=\sum_{h\in NK}\mathcal{T}^{l*}_{hf}\mathcal{T}^J_{hi}+\sum_{h\in K}\mathcal{T}^{J*}_{hf}\mathcal{T}^J_{hi}.
        \eeqn
        The non-critical block of the scattering matrix $\mathcal{T}^J$ --- $\mathcal{T}\big|_{NK\times NK}$--- can be safely diagonalized through a unitary matrix $U$. We then get,
        \beqa
                \Im\,U\mathcal{T}\big|_{NK\times NK}U^{-1}&=&U\mathcal{T}^{J\dagger}\big|_{NK\times NK}U^{-1}U\mathcal{T}\big|_{NK\times NK}U^{-1}\nonumber\\
                                                                     &&+ U\mathcal{T}^{J\dagger}\big|_{NK\times K}U^{-1}U\mathcal{T}\big|_{K\times NK}U^{-1}.
        \eeqan
        Denoting $\mathcal{T}\big|_{NK\times NK}$ by $\tilde{\mathcal{T}}$,
        \beq
                \Im\,\tilde{\mathcal{T}}_{ii} = |\tilde{\mathcal{T}}_{ii}|^2 + \sum_{h\in K}\left| \left( \mathcal{T}^J\big|_{K\times NK}U^{-1}\right)_{ih}\right|^2.
        \eeqn
    We then obtain a modified version of ~\leqn{Eq: circle}
    \begin{align}
        \left(\mathrm{Re}\tilde{\mathcal{T}}_{ii}\right)^2 + \left(\mathrm{Im}\tilde{\mathcal{T}}_{ii} - \frac{1}{2}\right)^2 = \frac{1}{4} - R^2
    \end{align}
with a small correction $R$ such that $R^2=\sum_{h\in K}\left| \left( \mathcal{T}^J|_{K\times NK}U^{-1}\right)_{ih}\right|^2\ge0$. Partial diagonalization then makes the radius of the Argand circle shrink from $1/2$ to $\sqrt{1/4-R^2}$. One can retrieve the original Argand circle by defining
\begin{align}
    \tilde{x} = \sqrt{\left(\mathrm{Re}\tilde{M}_{ii}\right)^2 - R^2} \quad\quad \tilde{y} = \mathrm{Im}\tilde{M}_{ii}.
\end{align}
The procedure and identities shown in the previous sections remain unchanged if $\tilde{x}$ and $\tilde{y}$ are used instead of $\mathrm{Re} \tilde{\mathcal{T}}_{ii}$ and $\mathrm{Im}\tilde{\mathcal{T}}_{ii}$.

\section{$\sqrt{s}\to\infty$ S-wave unitarity matrix}
\label{App: sinf}

$\M^0_{CP+}=$
\beq
\begin{sideways}
\begin{minipage}{\textheight}
\scalebox{0.58}{$
\left(
\begin{array}{ccccccccccc}
 -\frac{3}{8} \left(\text{g1}^2+\text{g2}^2\right) & \frac{1}{8} \left(-\text{g1}^2-\text{g2}^2\right) & \frac{1}{8} \left(\text{g1}^2+\text{g2}^2-4 \lambda ^2\right) & \frac{1}{8} \left(\text{g1}^2+\text{g2}^2-4 \lambda ^2\right) & 0 & 0 & -\frac{\lambda ^2}{2} & -\frac{\lambda ^2}{2} & -\frac{\text{g1}^2+\text{g2}^2}{4 \sqrt{2}} & \frac{\text{g1}^2-\text{g2}^2}{4 \sqrt{2}} & 0 \\
 \frac{1}{8} \left(-\text{g1}^2-\text{g2}^2\right) & -\frac{3}{8} \left(\text{g1}^2+\text{g2}^2\right) & \frac{1}{8} \left(\text{g1}^2+\text{g2}^2-4 \lambda ^2\right) & \frac{1}{8} \left(\text{g1}^2+\text{g2}^2-4 \lambda ^2\right) & 0 & 0 & -\frac{\lambda ^2}{2} & -\frac{\lambda ^2}{2} & -\frac{\text{g1}^2+\text{g2}^2}{4 \sqrt{2}} & \frac{\text{g1}^2-\text{g2}^2}{4 \sqrt{2}} & 0 \\
 \frac{1}{8} \left(\text{g1}^2+\text{g2}^2-4 \lambda ^2\right) & \frac{1}{8} \left(\text{g1}^2+\text{g2}^2-4 \lambda ^2\right) & -\frac{3}{8} \left(\text{g1}^2+\text{g2}^2\right) & \frac{1}{8} \left(-\text{g1}^2-\text{g2}^2\right) & 0 & 0 & -\frac{\lambda ^2}{2} & -\frac{\lambda ^2}{2} & \frac{\text{g1}^2-\text{g2}^2}{4 \sqrt{2}} & -\frac{\text{g1}^2+\text{g2}^2}{4 \sqrt{2}} & 0 \\
 \frac{1}{8} \left(\text{g1}^2+\text{g2}^2-4 \lambda ^2\right) & \frac{1}{8} \left(\text{g1}^2+\text{g2}^2-4 \lambda ^2\right) & \frac{1}{8} \left(-\text{g1}^2-\text{g2}^2\right) & -\frac{3}{8} \left(\text{g1}^2+\text{g2}^2\right) & 0 & 0 & -\frac{\lambda ^2}{2} & -\frac{\lambda ^2}{2} & \frac{\text{g1}^2-\text{g2}^2}{4 \sqrt{2}} & -\frac{\text{g1}^2+\text{g2}^2}{4 \sqrt{2}} & 0 \\
 0 & 0 & 0 & 0 & \frac{1}{4} \left(\text{g1}^2+\text{g2}^2-4 \lambda ^2\right) & 0 & \frac{\kappa  \lambda }{\sqrt{2}} & -\frac{\kappa  \lambda }{\sqrt{2}} & 0 & 0 & \frac{1}{4} \left(2 \lambda ^2-\text{g2}^2\right) \\
 0 & 0 & 0 & 0 & 0 & \frac{1}{4} \left(\text{g1}^2+\text{g2}^2-4 \lambda ^2\right) & -\frac{\kappa  \lambda }{\sqrt{2}} & \frac{\kappa  \lambda }{\sqrt{2}} & 0 & 0 & \frac{1}{4} \left(\text{g2}^2-2 \lambda ^2\right) \\
 -\frac{\lambda ^2}{2} & -\frac{\lambda ^2}{2} & -\frac{\lambda ^2}{2} & -\frac{\lambda ^2}{2} & \frac{\kappa  \lambda }{\sqrt{2}} & -\frac{\kappa  \lambda }{\sqrt{2}} & -3 \kappa ^2 & -\kappa ^2 & -\frac{\lambda ^2}{\sqrt{2}} & -\frac{\lambda ^2}{\sqrt{2}} & -\frac{\kappa  \lambda }{\sqrt{2}} \\
 -\frac{\lambda ^2}{2} & -\frac{\lambda ^2}{2} & -\frac{\lambda ^2}{2} & -\frac{\lambda ^2}{2} & -\frac{\kappa  \lambda }{\sqrt{2}} & \frac{\kappa  \lambda }{\sqrt{2}} & -\kappa ^2 & -3 \kappa ^2 & -\frac{\lambda ^2}{\sqrt{2}} & -\frac{\lambda ^2}{\sqrt{2}} & \frac{\kappa  \lambda }{\sqrt{2}} \\
 -\frac{\text{g1}^2+\text{g2}^2}{4 \sqrt{2}} & -\frac{\text{g1}^2+\text{g2}^2}{4 \sqrt{2}} & \frac{\text{g1}^2-\text{g2}^2}{4 \sqrt{2}} & \frac{\text{g1}^2-\text{g2}^2}{4 \sqrt{2}} & 0 & 0 & -\frac{\lambda ^2}{\sqrt{2}} & -\frac{\lambda ^2}{\sqrt{2}} & \frac{1}{2} \left(-\text{g1}^2-\text{g2}^2\right) & \frac{1}{4} \left(\text{g1}^2+\text{g2}^2-4 \lambda ^2\right) & 0 \\
 \frac{\text{g1}^2-\text{g2}^2}{4 \sqrt{2}} & \frac{\text{g1}^2-\text{g2}^2}{4 \sqrt{2}} & -\frac{\text{g1}^2+\text{g2}^2}{4 \sqrt{2}} & -\frac{\text{g1}^2+\text{g2}^2}{4 \sqrt{2}} & 0 & 0 & -\frac{\lambda ^2}{\sqrt{2}} & -\frac{\lambda ^2}{\sqrt{2}} & \frac{1}{4} \left(\text{g1}^2+\text{g2}^2-4 \lambda ^2\right) & \frac{1}{2} \left(-\text{g1}^2-\text{g2}^2\right) & 0 \\
 0 & 0 & 0 & 0 & \frac{1}{4} \left(2 \lambda ^2-\text{g2}^2\right) & \frac{1}{4} \left(\text{g2}^2-2 \lambda ^2\right) & -\frac{\kappa  \lambda }{\sqrt{2}} & \frac{\kappa  \lambda }{\sqrt{2}} & 0 & 0 & \frac{1}{4} \left(\text{g1}^2+\text{g2}^2-4 \lambda ^2\right)
\end{array}
\right)
$}
\end{minipage}
\end{sideways}
\eeqn

\section{Quartic couplings}
\beqa
Vssss(a,b,c,d)= \CR
\Bigg(-i{3\over4}(g_1^2+g_2^2)(&&U^S_{a,1}U^S_{b,1}U^S_{c,1}U^S_{d,1}+U^S_{a,2}U^S_{b,2}U^S_{c,2}U^S_{d,2})\CR
+i({1\over4}(g_1^2+g_2^2)-\lambda ^2)\big(&&U^S_{a,1}U^S_{b,1}U^S_{c,2}U^S_{d,2}+U^S_{a,1}U^S_{b,2}U^S_{c,1}U^S_{d,2}\CR
&&+U^S_{a,1}U^S_{b,2}U^S_{c,2}U^S_{d,1}+U^S_{a,2}U^S_{b,1}U^S_{c,1}U^S_{d,2}\CR
&&+U^S_{a,2}U^S_{b,1}U^S_{c,2}U^S_{d,1}+U^S_{a,2}U^S_{b,2}U^S_{c,1}U^S_{d,1}\big)\CR
-i \lambda ^2(&&U^S_{a,1}U^S_{b,1}U^S_{c,3}U^S_{d,3} + U^S_{a,1}U^S_{b,3}U^S_{c,1}U^S_{d,3} 
+ U^S_{a,1}U^S_{b,3}U^S_{c,3}U^S_{d,1} \CR
&&+ U^S_{a,3}U^S_{b,1}U^S_{c,1}U^S_{d,3} +U^S_{a,3}U^S_{b,1}U^S_{c,3}U^S_{d,1}+U^S_{a,3}U^S_{b,3}U^S_{c,1}U^S_{d,1} \CR
&&+U^S_{a,2}U^S_{b,2}U^S_{c,3}U^S_{d,3}+U^S_{a,2}U^S_{b,3}U^S_{c,2}U^S_{d,3}
+U^S_{a,2}U^S_{b,3}U^S_{c,3}U^S_{d,2}\CR
&&+U^S_{a,3}U^S_{b,2}U^S_{c,2}U^S_{d,3} + U^S_{a,3}U^S_{b,2}U^S_{c,3}U^S_{d,2}+U^S_{a,3}U^S_{b,3}U^S_{c,2}U^S_{d,2})\CR
-6i \kappa ^2(&&U^S_{a,3}U^S_{b,3}U^S_{c,3}U^S_{d,3})\CR
+i \lambda  \kappa (&&U^S_{a,1}U^S_{b,2}U^S_{c,3}U^S_{d,3}+U^S_{a,1}U^S_{b,3}U^S_{c,2}U^S_{d,3}\CR
&+&U^S_{a,1}U^S_{b,3}U^S_{c,3}U^S_{d,2}+U^S_{a,2}U^S_{b,1}U^S_{c,3}U^S_{d,3}\CR
&+&U^S_{a,2}U^S_{b,3}U^S_{c,1}U^S_{d,3}+U^S_{a,2}U^S_{b,3}U^S_{c,3}U^S_{d,1}\CR
&+&U^S_{a,3}U^S_{b,1}U^S_{c,2}U^S_{d,3}+U^S_{a,3}U^S_{b,1}U^S_{c,3}U^S_{d,2}\CR
&+&U^S_{a,3}U^S_{b,2}U^S_{c,1}U^S_{d,3}+U^S_{a,3}U^S_{b,2}U^S_{c,3}U^S_{d,1}\CR
&+&U^S_{a,3}U^S_{b,3}U^S_{c,1}U^S_{d,2}+U^S_{a,3}U^S_{b,3}U^S_{c,2}U^S_{d,1})\Bigg)
\eeqan
\beqa
Vsspp(a,b,\gamma,\delta)=\CR
\Bigg(-i/4(g_1^2+g_2^2)(&&U^S_{a,1}U^S_{b,1}U^P_{\gamma ,1}U^P_{\delta ,1}+U^S_{a,2}U^S_{b,2}U^P_{\gamma ,2}U^P_{\delta ,2})\CR
+i({1\over4}(g_1^2+g_2^2)-\lambda ^2)(&&U^S_{a,1}U^S_{b,1}U^P_{\gamma ,2}U^P_{\delta ,2}+U^S_{a,2}U^S_{b,2}U^P_{\gamma ,1}U^P_{\delta ,1})\CR
-i \lambda ^2(&&U^S_{a,1}U^S_{b,1}U^P_{\gamma ,3}U^P_{\delta ,3}+U^S_{a,3}U^S_{b,3}U^P_{\gamma ,1}U^P_{\delta ,1}\CR
&&+U^S_{a,2}U^S_{b,2}U^P_{\gamma ,3}U^P_{\delta ,3}+U^S_{a,3}U^S_{b,3}U^P_{\gamma ,2}U^P_{\delta ,2})\CR
-2i \kappa ^2(&&U^S_{a,3}U^S_{b,3}U^P_{\gamma ,3}U^P_{\delta ,3})\CR
-i \lambda  \kappa (&&(U^S_{a,1}U^S_{b,2}+U^S_{a,2}U^S_{b,1})U^P_{\gamma ,3}U^P_{\delta ,3}\CR
&&+U^S_{a,3}U^S_{b,3}(U^P_{\gamma ,1}U^P_{\delta ,2}+U^P_{\gamma ,2}U^P_{\delta ,1})\CR
&&-U^S_{a,1}U^S_{b,3}U^P_{\gamma ,2}U^P_{\delta ,3}-U^S_{a,3}U^S_{b,1}U^P_{\gamma ,2}U^P_{\delta ,3}\CR
&&-U^S_{a,1}U^S_{b,3}U^P_{\gamma ,3}U^P_{\delta ,2}-U^S_{a,3}U^S_{b,1}U^P_{\gamma ,3}U^P_{\delta ,2}\CR
&&-U^S_{a,2}U^S_{b,3}U^P_{\gamma ,1}U^P_{\delta ,3}-U^S_{a,3}U^S_{b,2}U^P_{\gamma ,1}U^P_{\delta ,3}\CR
&&-U^S_{a,2}U^S_{b,3}U^P_{\gamma ,3}U^P_{\delta ,1}-U^S_{a,3}U^S_{b,2}U^P_{\gamma ,3}U^P_{\delta ,1})\Bigg)\CR
\eeqan
\beqa
Vpppp(\alpha ,\beta ,\gamma ,\delta )=\CR
\Bigg(-3i/4(g_1^2+g_2^2)(&&U^P_{\alpha ,1}U^P_{\beta ,1}U^P_{\gamma ,1}U^P_{\delta ,1}+U^P_{\alpha ,2}U^P_{\beta ,2}U^P_{\gamma ,2}U^P_{\delta ,2})\CR
+i({1\over4}(g_1^2+g_2^2)-\lambda ^2)(&&U^P_{\alpha ,1}U^P_{\beta ,1}U^P_{\gamma ,2}U^P_{\delta ,2}+U^P_{\alpha ,1}U^P_{\beta ,2}U^P_{\gamma ,1}U^P_{\delta ,2}\CR
&&+U^P_{\alpha ,1}U^P_{\beta ,2}U^P_{\gamma ,2}U^P_{\delta ,1}+U^P_{\alpha ,2}U^P_{\beta ,1}U^P_{\gamma ,1}U^P_{\delta ,2}\CR
&&+U^P_{\alpha ,2}U^P_{\beta ,1}U^P_{\gamma ,2}U^P_{\delta ,1}+U^P_{\alpha ,2}U^P_{\beta ,2}U^P_{\gamma ,1}U^P_{\delta ,1})\CR
-i \lambda ^2(&&U^P_{\alpha ,1}U^P_{\beta ,1}U^P_{\gamma ,3}U^P_{\delta ,3}+U^P_{\alpha ,1}U^P_{\beta ,3}U^P_{\gamma ,1}U^P_{\delta ,3}\CR
&&+U^P_{\alpha ,1}U^P_{\beta ,3}U^P_{\gamma ,3}U^P_{\delta ,1}+U^P_{\alpha ,3}U^P_{\beta ,1}U^P_{\gamma ,1}U^P_{\delta ,3}\CR
&&+U^P_{\alpha ,3}U^P_{\beta ,1}U^P_{\gamma ,3}U^P_{\delta ,1}+U^P_{\alpha ,3}U^P_{\beta ,3}U^P_{\gamma ,1}U^P_{\delta ,1}\CR
&&+U^P_{\alpha ,2}U^P_{\beta ,2}U^P_{\gamma ,3}U^P_{\delta ,3}+U^P_{\alpha ,2}U^P_{\beta ,3}U^P_{\gamma ,2}U^P_{\delta ,3}\CR
&&+U^P_{\alpha ,2}U^P_{\beta ,3}U^P_{\gamma ,3}U^P_{\delta ,2}+U^P_{\alpha ,3}U^P_{\beta ,2}U^P_{\gamma ,2}U^P_{\delta ,3}\CR
&&+U^P_{\alpha ,3}U^P_{\beta ,2}U^P_{\gamma ,3}U^P_{\delta ,2}+U^P_{\alpha ,3}U^P_{\beta ,3}U^P_{\gamma ,2}U^P_{\delta ,2})\CR
-6i \kappa ^2(&&U^P_{\alpha ,3}U^P_{\beta ,3}U^P_{\gamma ,3}U^P_{\delta ,3})\CR
+i \lambda  \kappa (&&U^P_{\alpha ,1}U^P_{\beta ,2}U^P_{\gamma ,3}U^P_{\delta ,3}+U^P_{\alpha ,1}U^P_{\beta ,3}U^P_{\gamma ,2}U^P_{\delta ,3}\CR
&&+U^P_{\alpha ,1}U^P_{\beta ,3}U^P_{\gamma ,3}U^P_{\delta ,2}+U^P_{\alpha ,2}U^P_{\beta ,1}U^P_{\gamma ,3}U^P_{\delta ,3}\CR
&&+U^P_{\alpha ,2}U^P_{\beta ,3}U^P_{\gamma ,1}U^P_{\delta ,3}+U^P_{\alpha ,2}U^P_{\beta ,3}U^P_{\gamma ,3}U^P_{\delta ,1}\CR
&&+U^P_{\alpha ,3}U^P_{\beta ,1}U^P_{\gamma ,2}U^P_{\delta ,3}+U^P_{\alpha ,3}U^P_{\beta ,1}U^P_{\gamma ,3}U^P_{\delta ,2}\CR
&&+U^P_{\alpha ,3}U^P_{\beta ,2}U^P_{\gamma ,1}U^P_{\delta ,3}+U^P_{\alpha ,3}U^P_{\beta ,2}U^P_{\gamma ,3}U^P_{\delta ,1}\CR
&&+U^P_{\alpha ,3}U^P_{\beta ,3}U^P_{\gamma ,1}U^P_{\delta ,2}+U^P_{\alpha ,3}U^P_{\beta ,3}U^P_{\gamma ,2}U^P_{\delta ,1}))
\eeqan

\beqa
Vsscc(a,b)=\Bigg(-(i/4)g_2^2(&&U^S_{a,2}U^S_{b,2}+U^S_{a,1}U^S_{b,1})\CR
-i(g_2^2/4-\lambda ^2/2)(&&U^S_{a,1}U^S_{b,2}+U^S_{a,2}U^S_{b,1})\sin2 \beta\CR
-i ({1\over4})g_1^2(&&U^S_{a,2}U^S_{b,2}-U^S_{a,1}U^S_{b,1})\cos2\beta\CR
-i \lambda (\lambda  +\kappa  \sin2\beta)&&U^S_{a,3}U^S_{b,3}\Bigg)
\eeqan
\beqa
Vwwss(a,b)=\Big(-i({1\over4})g_2^2(&&U^S_{a,2}U^S_{b,2}+U^S_{a,1}U^S_{b,1})\CR
+i(g_2^2/4-\lambda ^2/2)(&&U^S_{a,1}U^S_{b,2}+U^S_{a,2}U^S_{b,1})\sin2\beta\CR
+i ({1\over4})g_1^2(&&U^S_{a,2}U^S_{b,2}-U^S_{a,1}U^S_{b,1})\cos2\beta\CR
-i \lambda (\lambda  -\kappa  \sin2\beta)&&U^S_{a,3}U^S_{b,3}\Big)
\eeqan
\beqa
Vppcc(\alpha ,\beta)=\Bigg(-(i/4)g_2^2(&&U^P_{\alpha ,2}U^P_{\beta ,2}+U^P_{\alpha ,1}U^P_{\beta ,1})\CR
+i(g_2^2/4-\lambda ^2/2)(&&U^P_{\alpha ,1}U^P_{\beta ,2}+U^P_{\alpha ,2}U^P_{\beta ,1})\sin2\beta\CR
-i ({1\over4})g_1^2(&&U^P_{\alpha ,2}U^P_{\beta ,2}-U^P_{\alpha ,1}U^P_{\beta ,1})\cos2\beta\CR
-i \lambda (\lambda  -\kappa  \sin2\beta)&&U^P_{\alpha ,3}U^P_{\beta ,3}\Bigg)
\eeqan
\beqa
Vwwpp(\alpha ,\beta)=\Bigg(-(i/4)g_2^2(&&U^P_{\alpha ,2}U^P_{\beta ,2}+U^P_{\alpha ,1}U^P_{\beta ,1})\CR
-i(g_2^2/4-\lambda ^2/2)(&&U^P_{\alpha ,1}U^P_{\beta ,2}+U^P_{\alpha ,2}U^P_{\beta ,1})\sin2\beta\CR
+i ({1\over4})g_1^2(&&U^P_{\alpha ,2}U^P_{\beta ,2}-U^P_{\alpha ,1}U^P_{\beta ,1})\cos2\beta\CR
-i \lambda (\lambda  +\kappa  \sin2\beta)&&U^P_{\alpha ,3}U^P_{\beta ,3}\Bigg)
\eeqan
\beqa
Vcccc&=&-i (\lambda ^2\sin[2{\beta}]^2+1/2(g_1^2+g_2^2)\cos[2{\beta}]^2)\CR
Vwwww&=&-i (\lambda ^2\sin[2{\beta}]^2+1/2(g_1^2+g_2^2)\cos[2{\beta}]^2)\CR
Vc^+c^-w^+w^- &=& i/4((g_1^2+g_2^2)\cos[4 {\beta}]-4\lambda ^2\cos[2{\beta}]^2)\CR
Vc^+w^-sp(a,\alpha)&=&\Big(\lambda  \kappa  U^S_{a,3}U^P_{\alpha ,3}\CR
&&-{1\over4}(g_2^2-2\lambda ^2)(U^S_{a,2}U^P_{\alpha ,1}-U^S_{a,1}U^P_{\alpha ,2}))\CR
Vc^-w^+sp(a,\alpha)&=&-\Big(\lambda  \kappa  U^S_{a,3}U^P_{\alpha ,3}\CR
&&-{1\over4}(g_2^2-2\lambda ^2)(U^S_{a,2}U^P_{\alpha ,1}-U^S_{a,1}U^P_{\alpha ,2})\Big)\CR
Vccww&=&-i/2(g_1^2+g_2^2-2 \lambda ^2)\sin[2{\beta}]^2\CR
Vsscw(a,b)&=&-i\Big((g_1^2/4)\sin2{\beta}(U^S_{a,1}U^S_{b,1}-U^S_{a,2}U^S_{b,2})\CR
&&+\lambda  \kappa  \cos[2{\beta}]U^S_{a,3}U^S_{b,3}\CR
&&+1/2(g_2^2/2-\lambda ^2)\cos[2{\beta}](U^S_{a,2}U^S_{b,1}+U^S_{a,1}U^S_{b,2})\Big)
\eeqan
\beqa
Vppcw(\alpha ,\beta)=-i\Big((g_1^2/4)\sin2{\beta}(&&U^P_{\alpha ,1}U^P_{\beta ,1}-U^P_{\alpha ,2}U^P_{\beta ,2})\CR
-\lambda  \kappa  \cos2{\beta}&&U^P_{\alpha ,3}U^P_{\beta ,3}\CR
-1/2(g_2^2/2-\lambda ^2)\cos2{\beta}(&&U^P_{\alpha ,2}U^P_{\beta ,1}+U^P_{\alpha ,1}U^P_{\beta ,2})\Big)
\eeqan
\beqa
Vcccw&=&i/4(g_1^2+g_2^2-2\lambda ^2)\sin[4 {\beta}]\CR
Vwwwc&=&-i/4(g_1^2+g_2^2-2\lambda ^2)\sin[4 {\beta}]\CR
\eeqan

\section{Tri-linear couplings}
\beqa
Vsss(a,b,c)=\Bigg(-(3i/(2\sqrt{2}))(g_1^2+g_2^2)(&&v_d  U^S_{a,1}U^S_{b,1}U^S_{c,1}+v_u  U^S_{a,2}U^S_{b,2}U^S_{c,2})\CR+i((g_1^2+g_2^2)/(2\sqrt{2})-\sqrt{2}\lambda ^2)v_d  (&& U^S_{a,1}U^S_{b,2}U^S_{c,2}+U^S_{a,2}U^S_{b,1}U^S_{c,2}+U^S_{a,2}U^S_{b,2}U^S_{c,1})\CR
+i((g_1^2+g_2^2)/(2\sqrt{2})-\sqrt{2}\lambda ^2)v_u (&&U^S_{a,1}U^S_{b,1}U^S_{c,2}+U^S_{a,1}U^S_{b,2}U^S_{c,1}+U^S_{a,2}U^S_{b,1}U^S_{c,1})\CR
+\sqrt{2}i(\lambda  \kappa  v_u  - \lambda ^2 v_d )(&&U^S_{a,1}U^S_{b,3}U^S_{c,3}+U^S_{a,3}U^S_{b,1}U^S_{c,3}+U^S_{a,3}U^S_{b,3}U^S_{c,1})\CR
+\sqrt{2}i(\lambda  \kappa  v_d  - \lambda ^2 v_u )(&&U^S_{a,2}U^S_{b,3}U^S_{c,3}+U^S_{a,3}U^S_{b,2}U^S_{c,3}+U^S_{a,3}U^S_{b,3}U^S_{c,2})\CR
-\sqrt{2}i \lambda ^2 v_s (&&U^S_{a,1}U^S_{b,1}U^S_{c,3}+U^S_{a,1}U^S_{b,3}U^S_{c,1}+U^S_{a,3}U^S_{b,1}U^S_{c,1}\CR
&&+U^S_{a,2}U^S_{b,2}U^S_{c,3}+U^S_{a,2}U^S_{b,3}U^S_{c,2}+U^S_{a,3}U^S_{b,2}U^S_{c,2})\CR
+i \lambda (A_{\lambda}/\sqrt{2}+\sqrt{2}\kappa  v_s )(&&U^S_{a,1}U^S_{b,2}U^S_{c,3}+U^S_{a,1}U^S_{b,3}U^S_{c,2}+U^S_{a,2}U^S_{b,1}U^S_{c,3}\CR
&&+U^S_{a,2}U^S_{b,3}U^S_{c,1}+U^S_{a,3}U^S_{b,1}U^S_{c,2}+U^S_{a,3}U^S_{b,2}U^S_{c,1})\CR
+i(-\sqrt{2} \kappa  A_{\kappa} - 6\sqrt{2} \kappa ^2 v_s )(&&U^S_{a,3}U^S_{b,3}U^S_{c,3})\Bigg)
\eeqan
\beqa
Vspp(a,\gamma,\delta)=\Bigg(-(i/2)(g_1^2+g_2^2)/\sqrt{2}(&&v_d  U^S_{a,1}U^P_{\delta ,1}U^P_{\gamma ,1}+v_u  U^S_{a,2}U^P_{\delta ,2}U^P_{\gamma ,2})\CR
+i((g_1^2+g_2^2)/(2\sqrt{2})-\sqrt{2}\lambda ^2)(&&v_d  U^S_{a,1}U^P_{\delta ,2}U^P_{\gamma ,2}+v_u  U^S_{a,2}U^P_{\delta ,1}U^P_{\gamma ,1})\CR
-\sqrt{2}i(\lambda  \kappa  v_d  + \lambda ^2 v_u )( &&U^S_{a,2}U^P_{\delta ,3}U^P_{\gamma ,3})\CR
-\sqrt{2}i(\lambda  \kappa  v_u  + \lambda ^2 v_d )( &&U^S_{a,1}U^P_{\delta ,3}U^P_{\gamma ,3})\CR
-\sqrt{2}i \lambda ^2 v_s  U^S_{a,3}(&&U^P_{\delta ,1}U^P_{\gamma ,1}+U^P_{\delta ,2}U^P_{\gamma ,2})\CR
-i(2\sqrt{2} \kappa ^2 v_s  - \sqrt{2} \kappa  A_{\kappa})(&&U^S_{a,3}U^P_{\delta ,3}U^P_{\gamma ,3})\CR
+\sqrt{2}i \lambda  \kappa  U^S_{a,3}(&&v_d (U^P_{\delta ,2}U^P_{\gamma ,3}+U^P_{\delta ,3}U^P_{\gamma ,2})\CR
+ v_u (&&U^P_{\delta ,1}U^P_{\gamma ,3}+U^P_{\delta ,3}U^P_{\gamma ,1}))\CR
+i(\sqrt{2}\lambda  \kappa  v_s  - \lambda  A_{\lambda}/\sqrt{2})(&&U^S_{a,1}(U^P_{\delta ,2}U^P_{\gamma ,3}+U^P_{\delta ,3}U^P_{\gamma ,2})\CR
+U^S_{a,2}(&&U^P_{\delta ,1}U^P_{\gamma ,3}+U^P_{\delta ,3}U^P_{\gamma ,1}))\CR
-i(\sqrt{2}\lambda  \kappa  v_s  + \lambda  A_{\lambda}/\sqrt{2})U^S_{a,3}(&&U^P_{\delta ,1}U^P_{\gamma ,2}+U^P_{\delta ,2}U^P_{\gamma ,1})\Bigg)
\eeqan
\beqa
Vscc(a)=\Big(-i g_2 m_W (&&U^S_{a,1}\cos\beta + U^S_{a,2}\sin\beta)\CR
-(i/2)\sqrt{g_1^2+g_2^2}m_Z(&&U^S_{a,2}\sin\beta - U^S_{a,1}\cos\beta)\cos2\beta\CR
+(i \lambda ^2/\sqrt{2})(&&v_d  U^S_{a,2}+v_u  U^S_{a,1})\sin2\beta\CR
-i (\lambda  /\sqrt{2})(&&(2 \kappa  v_s  + A_{\lambda})\sin2 \beta + 2\lambda v_s )U^S_{a,3}\Big)
\eeqan
\beqa
Vwcs(a)=-i\Big(1/(2\sqrt{2})(&&U^S_{a,1}(g_1^2\sin[2 \beta]v_d +(g_2^2-2\lambda ^2)\cos[2\beta]v_u )\CR
+ && U^S_{a,2}(-g_1^2\sin2\beta v_u +(g_2^2-2\lambda ^2)\cos[2\beta]v_d ))\CR
+ &&U^S_{a,3}(\sqrt{2}\kappa  \mu  +\lambda  A_{\lambda}/\sqrt{2})\cos2\beta\Big)
\eeqan
\beqa
Vw^+c^-p(\alpha)=\Big(-\sqrt{2}(\kappa  \mu -\lambda  A_{\lambda}/2)&& U^P_{\alpha ,3}\CR
+(v_u /\sqrt{2})(\lambda ^2-g_2^2/2)&&U^P_{\alpha ,1}\CR
+(v_d /\sqrt{2})(\lambda ^2-g_2^2/2)&&U^P_{\alpha ,2}\Big)\CR
Vw^-c^+p(\alpha)=-\Big((-\sqrt{2}(\kappa  \mu -\lambda  A_{\lambda}/2) &&U^P_{\alpha ,3}\CR
+(v_u /\sqrt{2})(\lambda ^2-g_2^2/2)&&U^P_{\alpha ,1}\CR
+(v_d /\sqrt{2})(\lambda ^2-g_2^2/2)&&U^P_{\alpha ,2})\Big)
\eeqan
\beqa
Vsww(a)=-i\sqrt{2}\Big((\mu (\lambda -\sin[2\beta]\kappa )-\lambda  A_{\lambda} /2 \sin[2 \beta])&&U^S_{a,3}\CR
+{1\over4}((g_2^2-g_1^2\cos[2\beta])v_u  - (g_2^2-2\lambda ^2)v_d  \sin[2\beta])&&U^S_{a,2}\CR
+{1\over4}((g_2^2+g_1^2\cos[2\beta])v_d  - (g_2^2-2\lambda ^2)v_u  \sin[2\beta])&&U^S_{a,1}\Big)
\eeqan

\end{document}